\documentclass[letterpaper]{mn2e}
\usepackage{epsfig,natbib2,natbibmnfix}

\newcommand{\apj}{ApJ}

\newcommand{\mnras}{MNRAS}
\newcommand{\aap}{A\&A}

\newcommand{\apjl}{ApJL}

\newcommand{\nat}{Nature}

\def\ltsima{$\; \buildrel < \over \sim \;$}
\def\simlt{\lower.5ex\hbox{\ltsima}}
\def\gtsima{$\; \buildrel > \over \sim \;$}
\def\simgt{\lower.5ex\hbox{\gtsima}}
\newcommand\sgra{Sgr~A$^*$}
\newcommand\ledd{{L}_{\rm Edd}}

\def\msun{{\,{\rm M}_\odot}}
\newcommand\mbh{{\,{M}_{\rm bh}}}
\def\rsun{{\,R_\odot}}

\title[Star formation in a disc around \sgra]{Simulations of star
formation in a gaseous disc around \sgra -- a failed Active Galactic Nucleus}

\author[S.~Nayakshin, J.~Cuadra \& V.~Springel]
{\parbox{18cm}{Sergei Nayakshin$^1$\footnotemark[1], Jorge~Cuadra$^2$\footnotemark[2]
and Volker~Springel$^2$}\vspace{0.3cm}\\
$^1$ Department of Physics \& Astronomy, 
University of Leicester, Leicester, LE1 7RH, UK\\
$^2$ Max-Planck-Institut f\"{u}r Astrophysik, Karl-Schwarzschild-Stra\ss{}e 1,
85741 Garching bei M\"{u}nchen, Germany}

\begin{document}

\maketitle

\begin{abstract}
  We numerically model fragmentation of a gravitationally unstable gaseous
  disc under conditions that may be appropriate for the formation of the young
  massive stars observed in the central parsec of our Galaxy. In this study,
  we adopt a simple prescription with a locally constant cooling time.  We
  find that, for cooling times just short enough to induce disc fragmentation,
  stars form with a top-heavy Initial Mass Function (IMF), as observed in the
  Galactic Centre (GC). For shorter cooling times, the disc fragments much
  more vigorously, leading to lower average stellar masses. Thermal feedback
  associated with gas accretion onto protostars slows down disc
  fragmentation, as predicted by some analytical models. We also simulate the
  fragmentation of a gas stream on an eccentric orbit in a combined \sgra\
  plus stellar cusp gravitational potential.  The stream precesses,
  self-collides and forms stars with a top-heavy IMF.  None of our models
  produces large enough co-moving groups of stars that could account for the
  observed ``mini star cluster'' IRS13E in the GC.  In all of the
  gravitationally unstable disc models that we explored, star formation takes
  place too fast to allow any gas accretion onto the central super-massive
  black hole. While this can help to explain the quiescence of
  `failed AGN' such as \sgra, it poses a challenge for understanding the
  high gas accretion rates infered for many quasars.

\end{abstract}


\footnotetext[1]{E-mail: {\tt Sergei.Nayakshin@astro.le.ac.uk}}
\footnotetext[2]{Current address: JILA, University of Colorado,
Boulder, CO 80309-0440, USA}

\section{Introduction}
\label{intro}

There is no detailed understanding of how super-massive black holes (SMBHs)
gain their mass, except that it must be mainly through gaseous disc accretion
\citep{Yu02}. The standard accretion disc model \citep{Shakura73} shows that
accretion discs must be quite massive and cold at ``large'' (sub-parsec in
this context) distances from the SMBH.  Many theorists pointed out that these
discs should be unstable to self-gravity, and thus must form stars or giant
planets by gravitational fragmentation
\citep[e.g.,][]{Paczynski78,Kolykhalov80,Shlosman89,Collin99,Bertin99,Gammie01}.
This creates a dilemma for the field, as star formation might be a dynamical,
very fast process which may result in a complete transformation of the gas
into stars. The SMBH would then be starved of fuel. \cite{Goodman03,Sirko03}
have recently demonstrated that star formation cannot be quenched by stellar
feedback, unless one is prepared to grossly violate constraints that we have
from AGN spectra.

\cite{Goodman03} suggested that the feeding of SMBHs proceeds via direct
accretion of low angular momentum gas that settles in a small scale disc. Such
a disc would be too hot to allow star formation. The size of this
``no-star-formation region'' is $R \simlt 0.03$ pc, and is weakly dependent on
the SMBH mass. However, one realistically expects that an even larger amount
of gas settles at larger radii where it could collapse and form stars, or it
could accrete on the SMBH. Therefore the fate of these self-gravitating discs
is still very much interesting.

Current observations provide new impetus for theoretical work on the topic.
\cite{Wehner06,FerrareseEtal06} have shown that many galaxies host ``extremely
compact nuclei'' (ECN) -- star clusters located in the very central $\sim 5$
parsec of galaxies. Closer to home, in the centre of our Galaxy, about a
hundred massive young stars, mostly arranged into two stellar discs of $\sim
0.05$--0.5 parsec scale \citep{Levin03,Genzel03a,Paumard06}, circle the
super-massive black hole named \sgra. These stars are suspected to have been
formed in situ, as evidenced by the lack of stars outside the $\sim 0.5$
parsec region \citep{Paumard06,NS05}.  Several observational facts are
consistent with the hypothesis that these stars formed in a massive gaseous
disc \citep{Paumard06}. Had this star formation event not happened, \sgra\
could have been accreting gas from the gaseous disc even now. \sgra\ thus
failed to realise itself as an AGN because of the loss of gas to star
formation \citep{NC05}.

A number of important gaps in our understanding of young stars near \sgra\
remain. The counter clock-wise disc appears to host stars on more elliptical
orbits, and it is also geometrically thicker. The same disc also contains a
puzzling ``mini star cluster'', IRS13E, that consists of more than a dozen
stars and may be bound by an Intermediate Mass Black Hole (IMBH) of mass $\mbh
\simgt 10^3 \msun$ \citep{Paumard05,Schoedel05}.  The IMF of the observed
stars must be top-heavy according to several lines of evidence
\citep{NS05,Nayakshinetal06,Paumard06,AlexanderBA06}, which is not expected in
the most basic model of a fragmenting disc, as the Jeans mass there is
significantly sub-solar (e.g., Levin 2006, Nayakshin 2006, but see also Larson
2006).

In this paper, we discuss numerical simulations of star formation occurring in
a gaseous disc around \sgra. Given the numerical challenges in the problem, we
foresee that a reliable modelling of all the questions raised by the young
stars near \sgra\ will require an extended effort of constantly increasing
complexity. In this study we present numerical experiments with a locally
constant cooling time. This allows a convenient comparison with previous
analytical and numerical works that predicted the conditions when
fragmentation should take place. It might also form a basis for comparison
with future work.

Within our formalism with a locally constant cooling time, we find
that (i) circular and eccentric discs alike can gravitationally
fragment and form stars; (ii) the IMF of formed stars is a strong
function of cooling time, becoming top-heavy for marginally
star-forming discs; (iii) star formation feedback is indeed able to
slow down disc fragmentation, as suggested by several earlier
analytical papers, but it is not yet clear if it can alleviate the
fueling problem of the SMBHs; (iv) our simulations do form some
tightly bound binary stars but more populous systems (a la IRS13E) do
not survive long.

This paper is structured as follows. In Section~\ref{sec:methods}, we
discuss our simulation methodology and the basic conditions for disc
fragmentation. Section~\ref{sec:starformation} explains our sink
particle approach to treat star formation in more detail. We then
analyse the evolution after disc fragmentation and the IMF of the
formed stars in Section~\ref{sec:nofb}. The sensitivity of our results
to numerical resolution and feedback from stars is discussed in
Sections~\ref{sec:sens} and \ref{sec:imfandfb}, respectively. We then
examine elliptical orbits of a gaseous stream in
Section~\ref{sec:ecc}, and the question of the formation of mini
star-clusters in Section~\ref{sec:irs13}. Finally, we summarize and
conclude in Section~\ref{sec:conclusions}.

\section{Methods and disc fragmentation tests} \label{sec:methods}

We use the SPH/$N$-body code {\small GADGET-2} \citep{Springel01,
Springel05} to simulate the dynamics of stars and gas in the
(Newtonian) gravitational field of a point mass with $\mbh = 3.5
\times 10^6 \msun$. The code solves for the gas hydrodynamics via the
smoothed particle hydrodynamics (SPH) formalism.  The hydrodynamic
treatment of the gas includes adiabatic processes and artificial bulk
viscosity to resolve shocks. The stars are modelled as sink particles,
using the approach developed by \cite{SpringelEtal05}, modified in the
ways described below.

Table~1 lists some of the parameters and results of the simulations
presented in this paper. The tests described in this section are those
listed as S1--S5 (the ``S'' stands for ``standard''). The units of
length and mass used in the simulations are $R_{\rm U} = 1.2 \times
10^{17} \hbox{cm} \approx 0.04$~pc, which is equal to $1''$ at the
8.0~kpc distance to the GC, and $M_{\rm U}=3.5 \times 10^6 \msun$, the
mass of Sgr~A* \citep[e.g.,][]{Schoedel02}, respectively. The
dimensionless time unit is $t_{\rm U} = 1/\Omega(R_{\rm U})$, or about
60 years ($\Omega$ is defined just below Eqn.~\ref{q}). The masses of
SPH particles are typically around $0.01$ Solar masses (see Table~1).

\subsection{Gravitational collapse}\label{sec:collapse}

\cite{Toomre64} showed that a rotating disc is subject to gravitational instabilities when the
$Q$-parameter
\begin{equation}
Q = \frac{\Omega^2}{2\pi G \rho}
\label{q}
\end{equation}
becomes smaller than a critical value, which is close to unity. This form of
the equation assumes hydrostatic equilibrium to relate the disc sound speed to
its vertical thickness, $H$. $\Omega=(G\mbh/R^3)^{1/2}$ is the Keplerian
angular frequency, and $\rho$ is the vertically averaged disc density.
Gravitational collapse thus takes place when the gas density exceeds
\begin{equation}
\rho_{\rm BH} \equiv \frac{\Omega^2}{2\pi G}\;.
\label{rhobh}
\end{equation}
Ideally, numerical simulations should resolve the gravitational collapse down
to stellar scales. In practice, this is impossible for numerical reasons, and
instead collisionless ``sink particles'' are commonly introduced
\citep{Bate95} to model collapsing regions of very high density.  
To ensure that collapse is well under way when we introduce a sink
particle, we require that the gas density exceeds
\begin{equation}
\rho_{\rm crit} = \rho_0 + A_{\rm col} \rho_{\rm BH}\;,
\label{rhosink}
\end{equation}
where $\rho_0 = 5 \times 10^{-11}$ g~cm$^{-3}$, and $A_{\rm col}$ is a large
number (we tested values from a few to a few thousand). We in general find
that our results are not sensitive to the exact values of $\rho_0$ and $A_{\rm
col}$, provided they are large enough (Section~\ref{sec:acol}). Further
details of our sink particle treatment are given in Section~\ref{sec:sink}.

\subsection{Gas cooling}\label{sec:cooling}

We account for radiative cooling via a simple approach with a locally constant cooling
time. The cooling term in the energy equation is
\begin{equation}
\left(\frac{{\rm d}u}{{\rm d}t}\right)_{\rm cool} \; = \; - \; \frac{u}{t_{\rm
cool}(R)}\;,
\label{tcool}
\end{equation}
where $u$ is the internal energy of an SPH particle, and $R$ is the radial
location of the SPH particle, i.e. the distance to the SMBH.  We parameterize
the cooling time as a fixed fraction of the local dynamical time, 

\begin{equation}
t_{\rm cool}(R) = \beta \times t_{\rm dyn}(R)\;,
\end{equation}
where $t_{\rm dyn}= 1/\Omega$ and $\beta$ is a parameter of the simulations.
This simple model allows for a convenient validation of the simulation
methodology as such locally constant cooling time models were investigated in
detail in previous literature. \cite{Gammie01} has shown that self-gravitating
discs are bound to collapse if the cooling time, $t_{\rm cool}$, is shorter
than about $3/\Omega$ (i.e., $\beta < 3$). \cite{Rice05} presented a range of
runs that tested this fragmentation criterion in detail, and found that, for
the adiabatic index of $\gamma=5/3$ (as used throughout this paper), the disk
fragments as long as $\beta \le 6$.

Note that most of the simulations explored in this paper are performed
for circular initial gas orbits, and also for relatively small disc
(total gas) masses as compared with that of the SMBH. As will become
clear later, this implies that during the simulations gas particles
continue to follow nearly circular orbits, and thus their cooling time
is constant around their orbits. One exception to this is a test done
with eccentric initial gas orbits (Section~\ref{sec:ecc}).

\subsection{Fragmentation tests}\label{sec:fragm}

To compare our numerical approach with known results, we ran several tests
(S1--S5 in Table 1) in a setup reminiscent of that of \cite{Rice05}, who
modelled fragmentation of proto-stellar discs.  One key difference between
marginally self-gravitating proto-stellar and AGN discs is the relative disc
mass. Whereas the former become self-gravitating for a disc to central
object mass ratio of $M_{\rm disc}/M_* \sim 0.1-0.5$
\citep[e.g.,][]{Rafikov05}, where $M_*$ is the mass of the central star, the
latter become self-gravitating for a mass ratio as small as $M_{\rm
disc}/\mbh \simeq 0.003-0.01$
\citep[e.g.,][]{Gammie01,Goodman03,NC05,Levin06}. For marginally
self-gravitating discs, the ratio of disc height to radius, $H/R$, is of order of
the mass ratio, $H/R \sim M_{\rm disc}/\mbh$ \citep{Gammie01}. The disc
viscous time is
\begin{equation}
t_{\rm visc} \sim \alpha^{-1} \left(\frac{R}{H}\right)^{2} \Omega^{-1}\;,
\label{tvisc} 
\end{equation}
where $\alpha \sim 0.1-1$ is the effective viscosity parameter
\citep{Lin87,Gammie01} for self-gravitating discs. Thus, for $H/R \simlt
0.01$, the disc viscous time is some 4--6 orders of magnitude longer than the
disc dynamical time. This is in fact one of the reasons why the discs become
self-gravitating, as they are not able to heat up quickly enough via viscous
energy dissipation \citep{Shlosman89,Gammie01}.

\begin{figure}
\centerline{\psfig{file=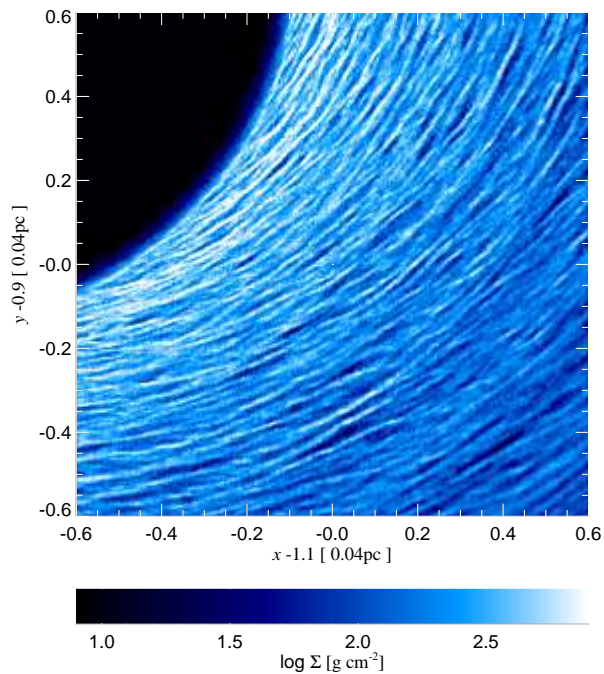,width=.48\textwidth,angle=0}}
\caption{Surface density of a region in a marginally stable disc from
the simulation S4 ($\beta = 4.5$) at time $t=50$. The map is centred on the
$(x,y) = (1.1, -0.9)$ location. Many small and somewhat dense gas
clumps can be seen in the figure. None of the clumps however truly
collapses, and most disappear on a dynamical time scale.}
\label{fig:fragm}
\end{figure}

Also, since the disc viscous time is many orders of magnitude longer than the
orbital time, as well as the total simulation time, we expect no radial
re-distribution of gas in the simulations, and indeed very little occurs. Our
models are essentially local, as emphasized by \cite{Nayakshin06a}, and hence
it suffices to simulate a small radial region of the disc. For definiteness,
we simulate a gaseous disc of mass $M_{\rm d} = 3 \times 10^4 \msun \approx
0.01 \mbh$ extending from $r_{\rm in}=1$ to $r_{\rm out}=4$, where $r$ is the
dimensionless distance from \sgra. 

The total initial number of SPH particles used in runs S1--S5 is $4\times
10^6$. Gravitational softening is adaptive, with a minimum gravitational
softening length of $3\times 10^{-4}$ for gas, and $0.001$ for sink
particles. The disk is in circular rotation around a SMBH with mass of $\mbh =
3.5\times 10^6\msun$.  The disc is extended vertically to a height of $H/R =
0.02$ in the initial conditions, which renders it stable to self-gravity. As
the gas settles into hydrostatic equilibrium, it heats up due to compressional
heating, and then cools according to equation~(\ref{tcool}).  We used five
different values of $\beta = t_{\rm cool} \Omega$, in the runs, namely $\beta
= 0.3$, $2$, $3$, $4.5$ and $6$ (see Table~1).

The discs in runs with $\beta=0.3, 2$ and 3 fragmented and formed stars, as
expected based on the results of \cite{Rice05}, but our tests with $\beta=4.5$
and $6$ did not. However, the run with $\beta=4.5$ and $6$ did fragment in the
sense of forming {\em transient} high density gas clumps, some of which can be
noted in Figure \ref{fig:fragm}. The maximum density in these clumps
fluctuated between a few to a few dozen times $\rho_{\rm BH}$, i.e. the clumps
were dense enough to be self-bound, but not dense enough for our sink particle
criterion, since $A_{\rm col}$ was set to 30 for these tests
(cf. Section~\ref{sec:collapse}).

\cite{Rice05} reported fragmentation for values of $\beta$ as large as 6. This
difference in the results is most likely due to our discs being much thinner
geometrically. The high density structures formed in our simulations, like
filaments and clumps, are much smaller in relative terms than they are in the
simulations of proto-stellar discs. This is not surprising as the most
unstable wavelength for the $Q\sim 1$ disc is of order of the disc vertical
scale height $H$ \citep{Toomre64}, and our discs are much thinner. Also, the
high density structures are much more numerous in AGN discs. As analytically
shown by \cite{Levin06}, self-gravitating clumps in such discs are quite
likely to collide with each other on a dynamical time scale. If clumps are
tightly bound they may agglomerate. For tests with larger values of $\beta$,
the clumps are marginally bound (as their maximum density exceeds $\rho_{\rm
BH}$ by a factor of a few only), and any interaction may unbind them. In
contrast, for high values of $H/R \sim M_{\rm disc}/M$, as appropriate for
proto-stellar discs, fewer clumps form, and the interactions are probably too
rare to cause clump destruction.

\section{Star formation}\label{sec:starformation}

\subsection{Two types of sink particles}\label{sec:sink}

To design a physically reasonable and numerically sound procedure to deal with
the gravitational collapse of cooling gas, we make use of well known results
from the field of star formation. Simulations of collapsing gas haloes show
that the first quasi-hydrostatic object to form has a characteristic size of
about 5 AU, and may therefore be called a ``first core''
\citep[e.g.,][]{Larson69,Masunaga98}. As this optically thick gas condensation
slowly cools, the core decreases in size while accreting mass from the
infalling envelope.  When the mass of the gas accumulated is of the order of
$0.05$--$0.1\msun$, H2 dissociation and hydrogen ionisation losses become
efficient coolants, and the core can collapse dynamically to much higher
densities and much smaller sizes \citep{Larson69,Silk77}. The relevant time
scale is of the order of $\simgt 10^3$ years, which is very short compared
with free fall times of typical molecular clouds in a galaxy. Creation of the
first collapsed objects is therefore nearly instantaneous in a ``normal'' star
forming environment, and their sizes are sufficiently small so that they can
be considered point masses compared with the cloud size of $\sim 10^{18}$~cm.

The gas densities in self-gravitating AGN discs are $\rho/m_{\rm p} \sim
10^8-10^{12}$~cm$^{-3}$, i.e. many orders of magnitude larger than in galactic
star forming regions, but smaller\footnote{Closer in to SMBHs, the tidal shear
density ($\rho_{\rm BH}$) becomes yet larger, and therefore there exists a
minimum radius in the disc inside of which star formation does not take place,
as proto-stars would be torn apart by tidal forces.} than the gas density of a
first core, $\sim$~few~$\times 10^{13}$~cm$^{-3}$. We assume that the initial
stages of collapse should be similar to ``normal'' star formation except that
it starts from higher initial gas densities. For \sgra, at the distance where
the young massive stars are located \citep{Paumard06}, the free-fall time in
the collapsing gas clouds, $t_{\rm ff} \sim 1/(G\rho)^{1/2}\simlt t_{\rm
dyn}$, is of the order of 100--1000 years. For a more massive AGN, the time
scales would be even shorter at comparable distances to the SMBH. The
life-times of the first cores in this environment are then longer than the
free-fall time, since the core collapse is dominated by the cooling time
scale, which is much longer than the $t_{\rm ff}$ estimated above. This
implies that these objects should actually be treated as having a finite
life-time in our simulations. Furthermore, their sizes are only an order of
magnitude smaller than that of the disc scale height, $H\sim 10^{15}$~cm
\citep{NC05}. Therefore, the first cores have a decent chance to merge with
one another \citep{Levin06}.  They could also be disrupted by a more massive
proto-star that passes by, form a disc and be subsequently accreted by the
proto-star.

In order to capture this complicated situation in a numerically
practical way, we introduce two kinds of sink particles. We call those
with mass $M\le M_{\rm core}= 0.1 \msun$ ``first cores''. Once the gas
reaches the critical density (eq.~\ref{rhosink}), we turn SPH
particles into sink particles individually, so the initial mass of a
first core particle is equal to one SPH particle mass. First cores
have geometrical sizes of $R_{\rm core}$, and can merge with one
another if they pass within a distance smaller than $2 f_m R_{\rm
core}$ of each other, where $f_m \ge 1$ is a parameter that mimics 
the effect of gravitational focusing
in mergers. We assume for simplicity that the lifetime of
the first cores is longer than the duration of the simulations. We
performed a few tests with finite core life times and found very
similar results; most of the merging of first cores happens very
quickly after they are created inside of a collapsing gas clump.

Sink particles with mass $M > M_{\rm core}$ are assumed to have collapsed to
stellar densities, thus we refer to them simply as ``stars''. Stars are
treated as point mass particles and thus cannot merge with one another, but
they are allowed to accrete the first cores if their separation is less than
$R_{\rm core}$. When the mass of a first core particle exceeds the critical
mass $M_{\rm core}$ as a result of mergers or gas accretion, we turn it into a
star particle.

\subsection{Accretion onto sink particles}

We use the Bondi-Hoyle formalism to calculate the accretion rate onto the sink
particles,
\begin{equation}
\dot M = 4 \pi \rho \frac{(G M)^2}{(\Delta v^2 + c_s^2)^{3/2}}\;,
\label{mbondi}
\end{equation}
where $M$ is the sink particle mass, and $\rho$, $c_s$ and $\Delta v$ are the
ambient gas density, sound speed, and the relative velocity between the
accretor and the gas, respectively. Once the accretion rate is computed, the
actual gas particles that are swallowed by the sink particles are chosen from
among the sink particle's neighbors via the stochastic SPH method of
\cite{SpringelEtal05}.

\subsection{Maximum accretion rates}\label{sec:medd}

As the gas density in these star forming discs is enormous compared to
``normal'' molecular clouds, the Bondi-Hoyle formula frequently yields
super-Eddington accretion rates \citep{Nayakshin06a}. Assuming that gas
liberates $G M_*/R_*$ of energy per unit mass, where $M_*$ and $R_*$ are the
mass and radius of the accretor, one estimates that the accretion luminosity is
\begin{equation}
L_{\rm accr} = \frac{GM_* \dot M}{R_*}\;,
\label{laccr}
\end{equation}
where $\dot M$ is the accretion rate. Setting this equal to the Eddington
luminosity, $\ledd = 4\pi G M_* m_p c/\sigma_T = 1.3 \times 10^{38}
(M_*/\msun)$~erg~s$^{-1}$, we obtain the Eddington accretion rate:
\begin{equation}
{\dot M}_{\rm Edd} = \frac{4\pi m_p R_* c}{\sigma_T}\;.
\label{medd}
\end{equation}
Note that this expression depends only on the size of the object, $R_*$. For
first cores, $R_* = R_{\rm core}$, which yields a very high accretion rate
limit of almost a Solar mass per year for $R_{\rm core} = 10^{14}$ cm. For
stars, we use the observational results of \cite{1991Ap&SS.181..313D,
1998ARep...42..793G}:
\begin{eqnarray}
\frac{R}{\rsun} & = & 1.09 \left(\frac{M}{\msun}\right)^{0.969}\;\;{\rm for}\;\;   M < 1.52 \msun, \\
\frac{R}{\rsun} & = & 1.29 \left(\frac{M}{\msun}\right)^{0.6035} \;\;{\rm for}\;\;  M > 1.52 \msun .
\end{eqnarray}
For $R=\rsun$, this yields
an Eddington accretion rate limit of $\sim 5 \times 10^{-4} \msun$~year$^{-1}$.

\section{Disc evolution after fragmentation}\label{sec:nofb}

We use the tests with cooling parameter $\beta=2$ and $3$, briefly
described in Section~\ref{sec:fragm}, to point out some general trends
seen in our simulations. These runs were continued far longer than was
needed to form the first gravitationally bound clumps. Sink particles
were introduced as described in Sections~\ref{sec:fragm} and
\ref{sec:sink}. The collapse parameter $A_{\rm col}$ was set to 30
(see equation \ref{rhosink}), and the gravitational focusing parameter
is $f_m = 3$.

\begin{figure*}
\begin{minipage}[b]{.48\textwidth}
\centerline{\psfig{file=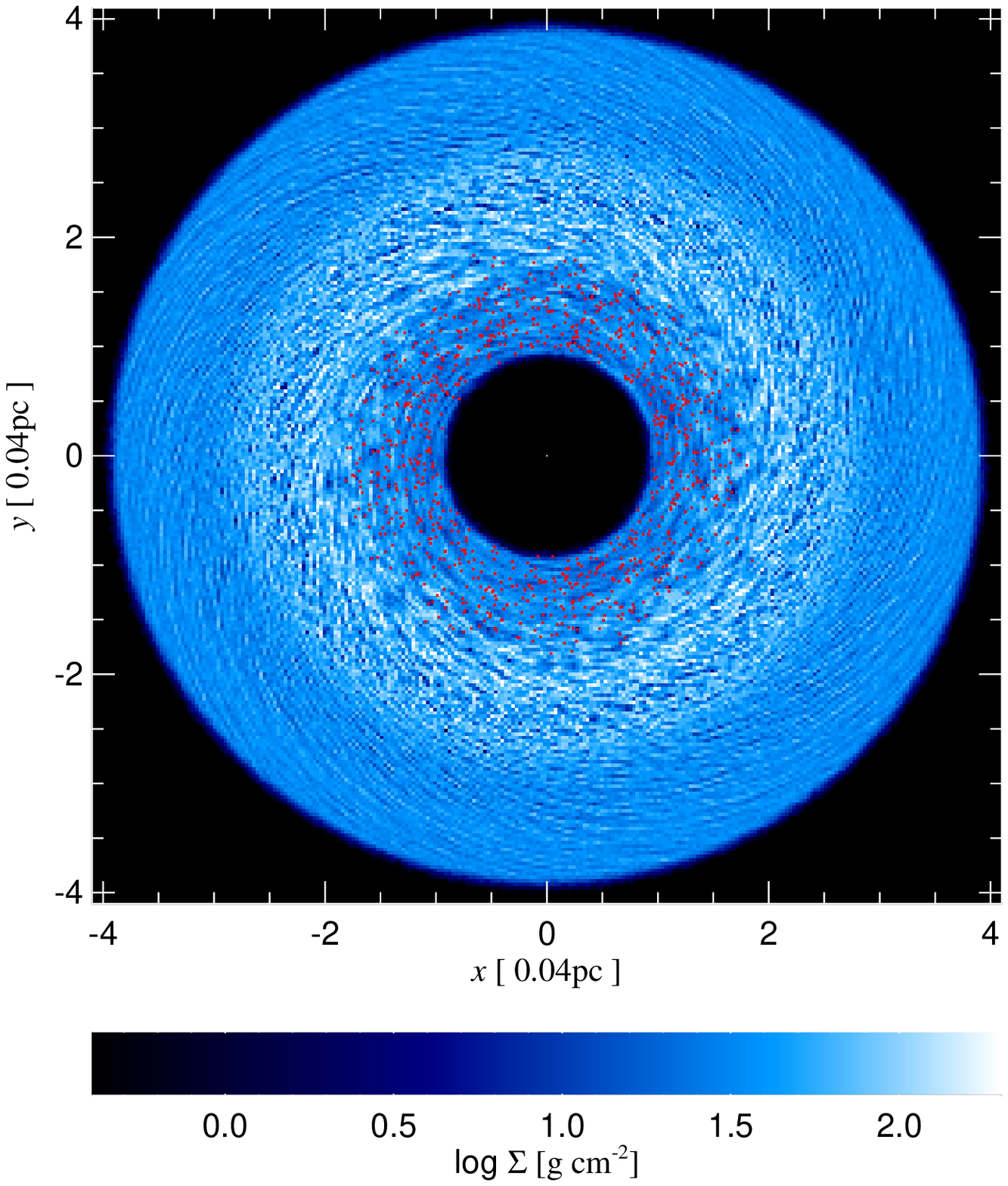,width=1.\textwidth,angle=0}}
\end{minipage}
\begin{minipage}[b]{.48\textwidth}
\centerline{\psfig{file=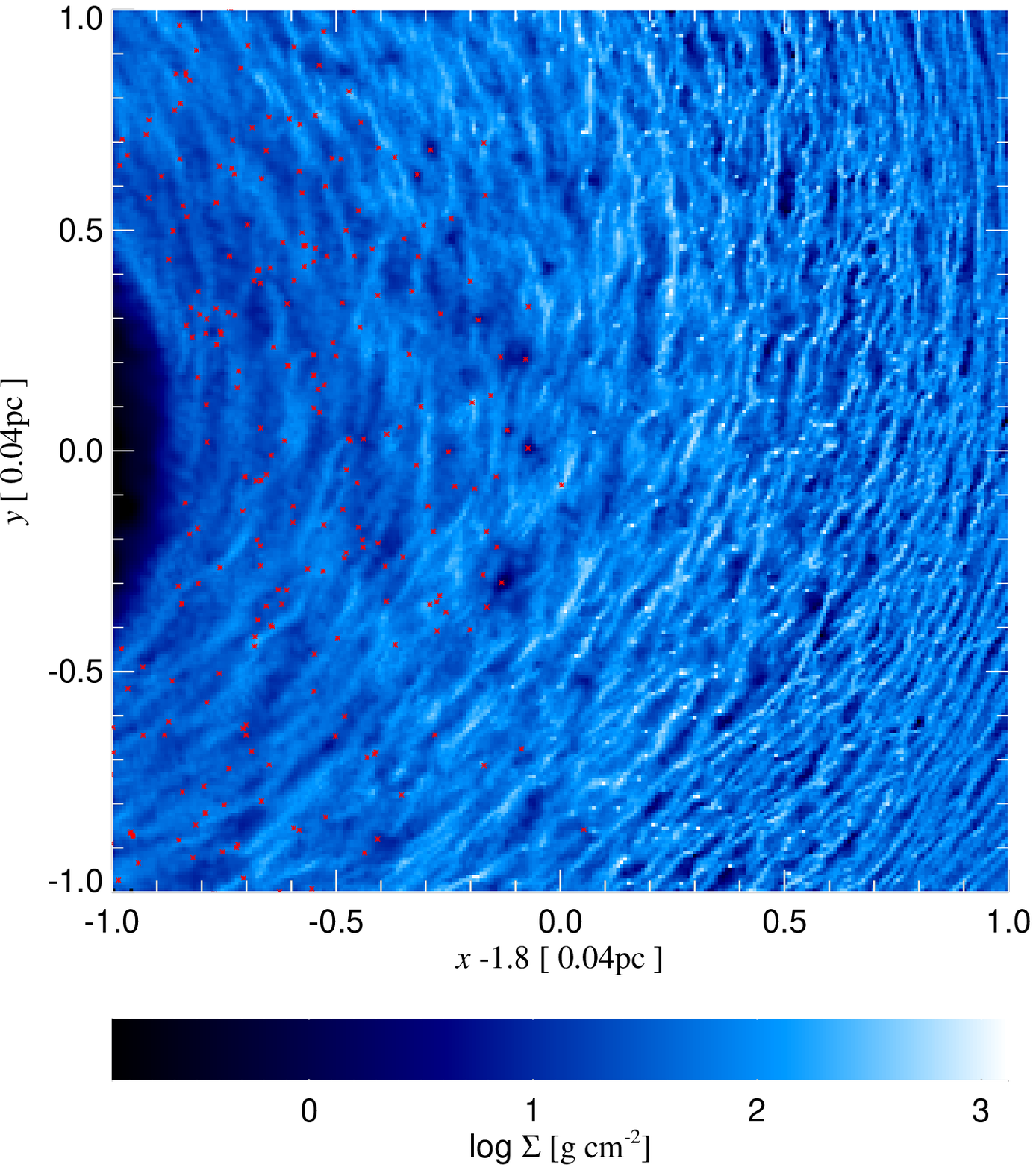,width=1.\textwidth,angle=0}}
\end{minipage}
\caption{Snapshot of the disc column density at time $t=75$ for test
S2 ($\beta=2$). The gas is rotating clockwise in this and all the
other figures in this paper. The left hand panel shows the full
simulation domain, whereas the right hand one zooms in on a region of
the disc centred at $x=1.8$. Stars with masses greater than $3 \msun$
are plotted as red asterisks.}
\label{fig:fig1}
\end{figure*}

Figure~\ref{fig:fig1} shows a column density map of a snapshot made at
dimensionless time $t=75$ for run S2 ($\beta=2$). The left panel of the figure
shows the whole disc, whereas the right panel zooms in on a smaller region
centred on $x=1.8$. Red asterisks show stars more massive that 3$\msun$. We
do not show lighter stars and first cores for clarity in the figure. Since gas
dynamical and cooling times are shorter for smaller radii, high density
regions develop faster at smaller radii. 

The sequence of events is as follows. The innermost edge of the disc forms
collapsed haloes in which sink particles are introduced. These sink particles
grow by mergers with other sink particles and gas accretion. With time, the
gaseous disc is depleted while the stellar disc grows.

Looking at larger $R$ in Fig.~\ref{fig:fig1}, we observe an intermediate epoch
in star formation, where bound high density clumps have formed but there are
no stars more massive than 3$\msun$ yet. This is because some of these clumps
have not yet satisfied the fragmentation criterion, and others satisfied it
only recently, thus containing only first cores and low-mass stars. Finally,
at the largest radii, we see dense filaments (spiral structure) but no well
defined bound clumps.

\begin{figure*}
\begin{minipage}[b]{.48\textwidth}
\centerline{\psfig{file=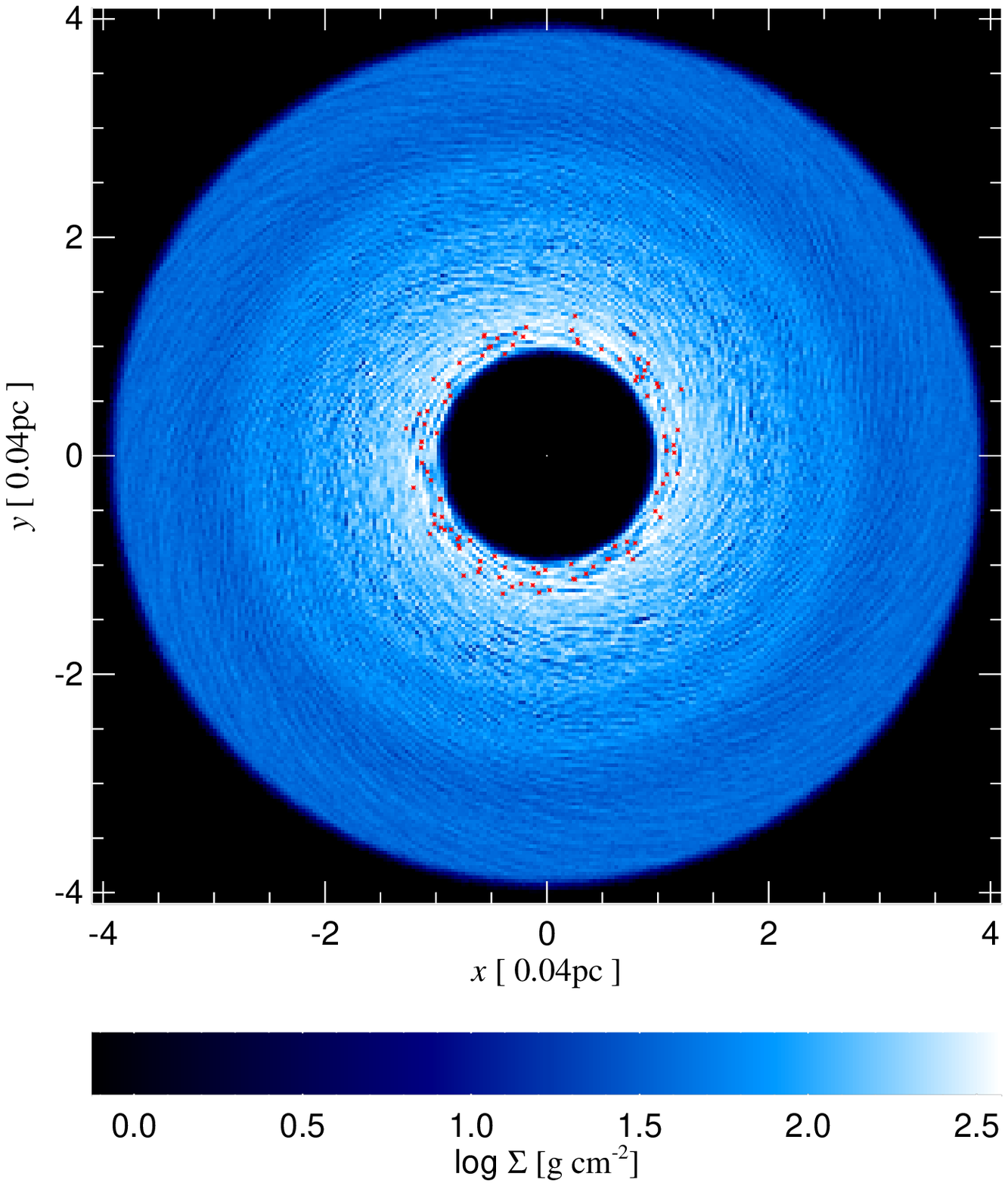,width=1.\textwidth,angle=0}}
\end{minipage}
\begin{minipage}[b]{.48\textwidth}
\centerline{\psfig{file=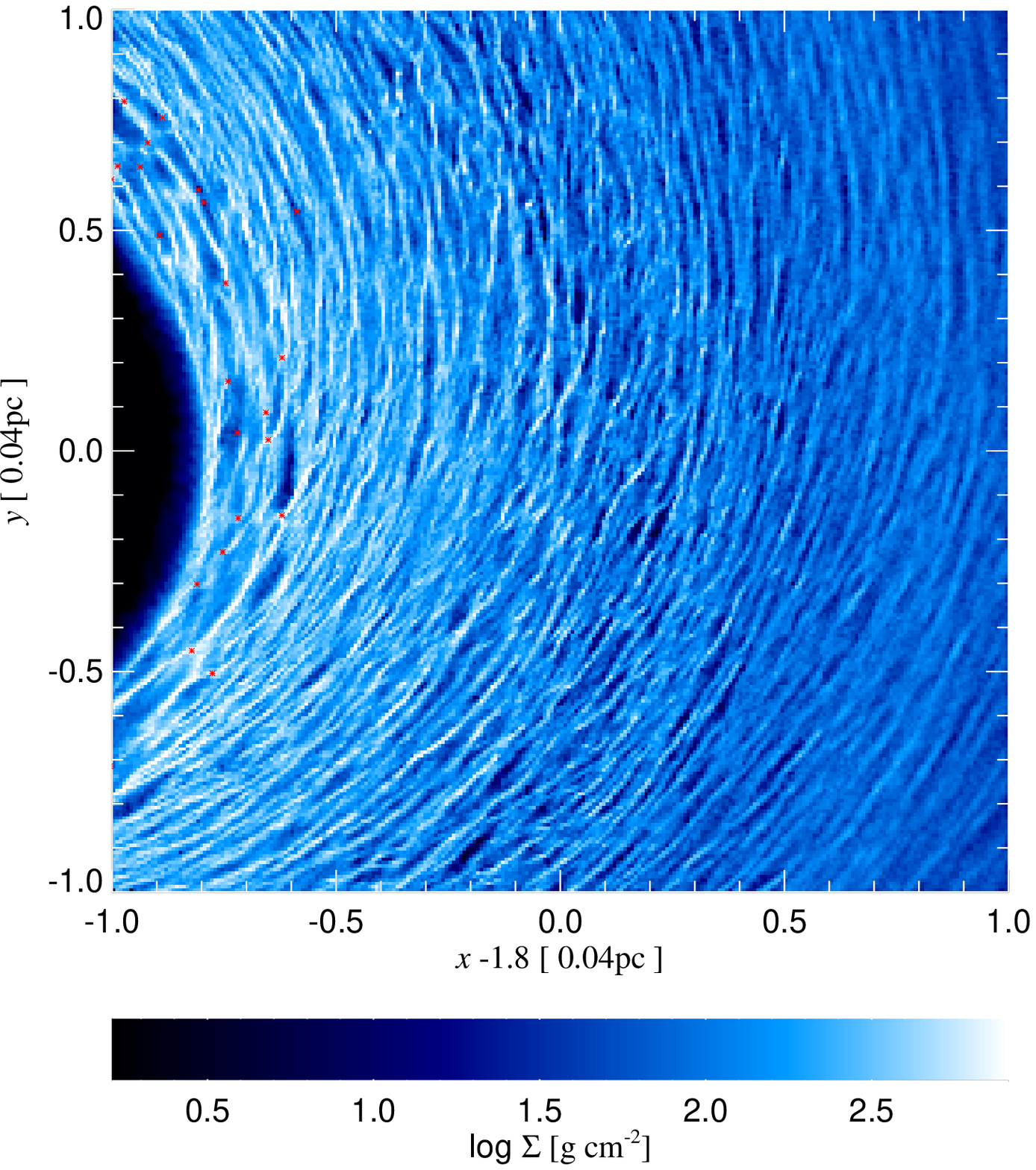,width=1.\textwidth,angle=0}}
\end{minipage}
\caption{Same as Fig.~\ref{fig:fig1} but for test S3 ($\beta=3$),
  i.e.~for a longer cooling time. Note the smaller number of stars and
  high density clumps in this test compared with Fig.~\ref{fig:fig1},
  where the cooling time is shorter.}
\label{fig:fig2}
\end{figure*}

Figure~\ref{fig:fig2} is identical to Fig.~\ref{fig:fig1}, but shows test S3
(i.e. $\beta=3$ instead of $\beta=2$). Comparing the two runs, it is
apparent that fragmentation of the more gradually cooling disc (S3) is
naturally much slower than in the test S2. There are fewer stars in the
disc, and there are fewer high density clumps. As we shall see later, within
the given model, this allows the stars to grow to larger masses, resulting in
a more top-heavy mass function. 

\begin{figure*}
\begin{minipage}[b]{.48\textwidth}
\centerline{\psfig{file=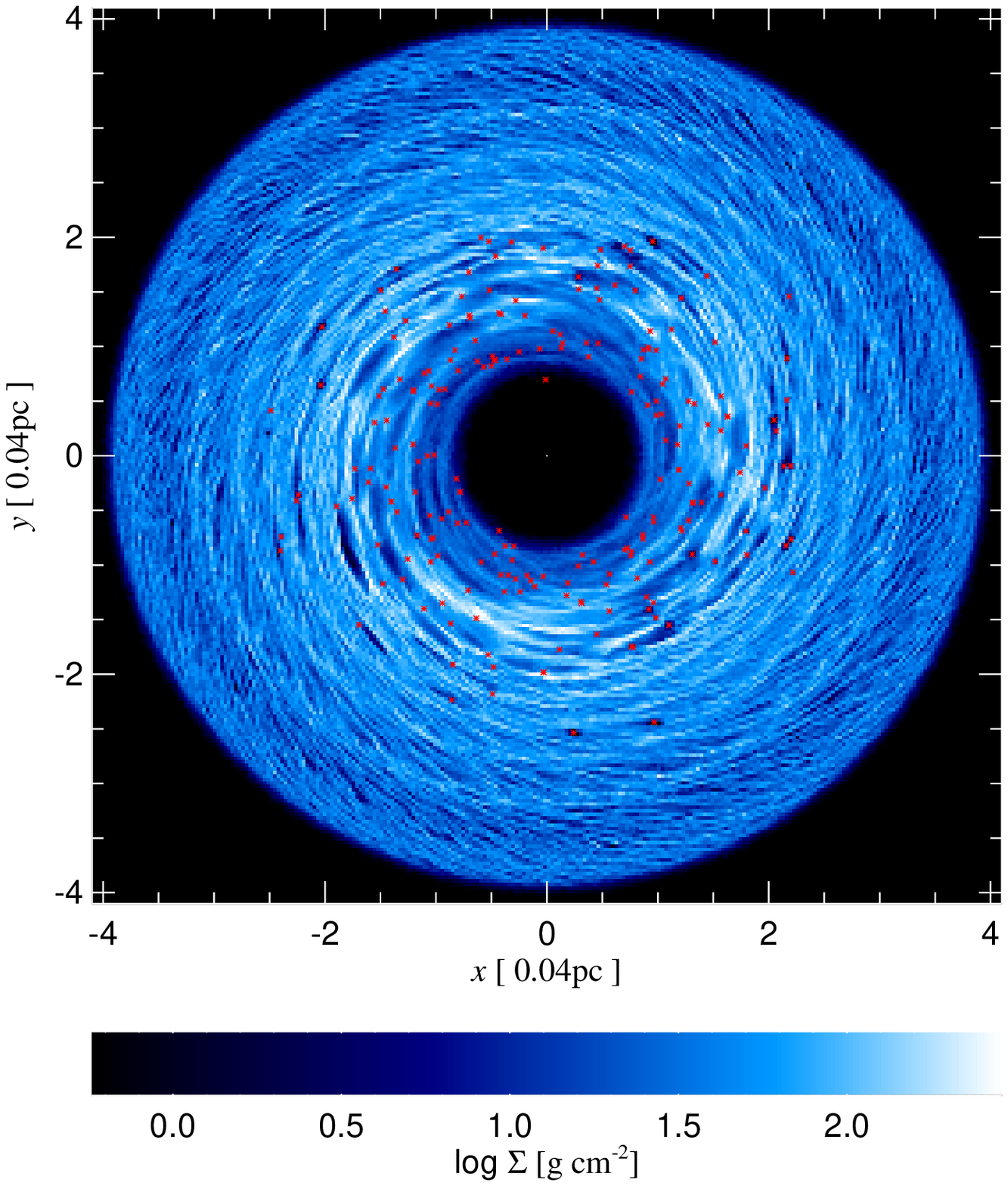,width=0.99\textwidth,angle=0}}
\end{minipage}
\begin{minipage}[b]{.48\textwidth}
\centerline{\psfig{file=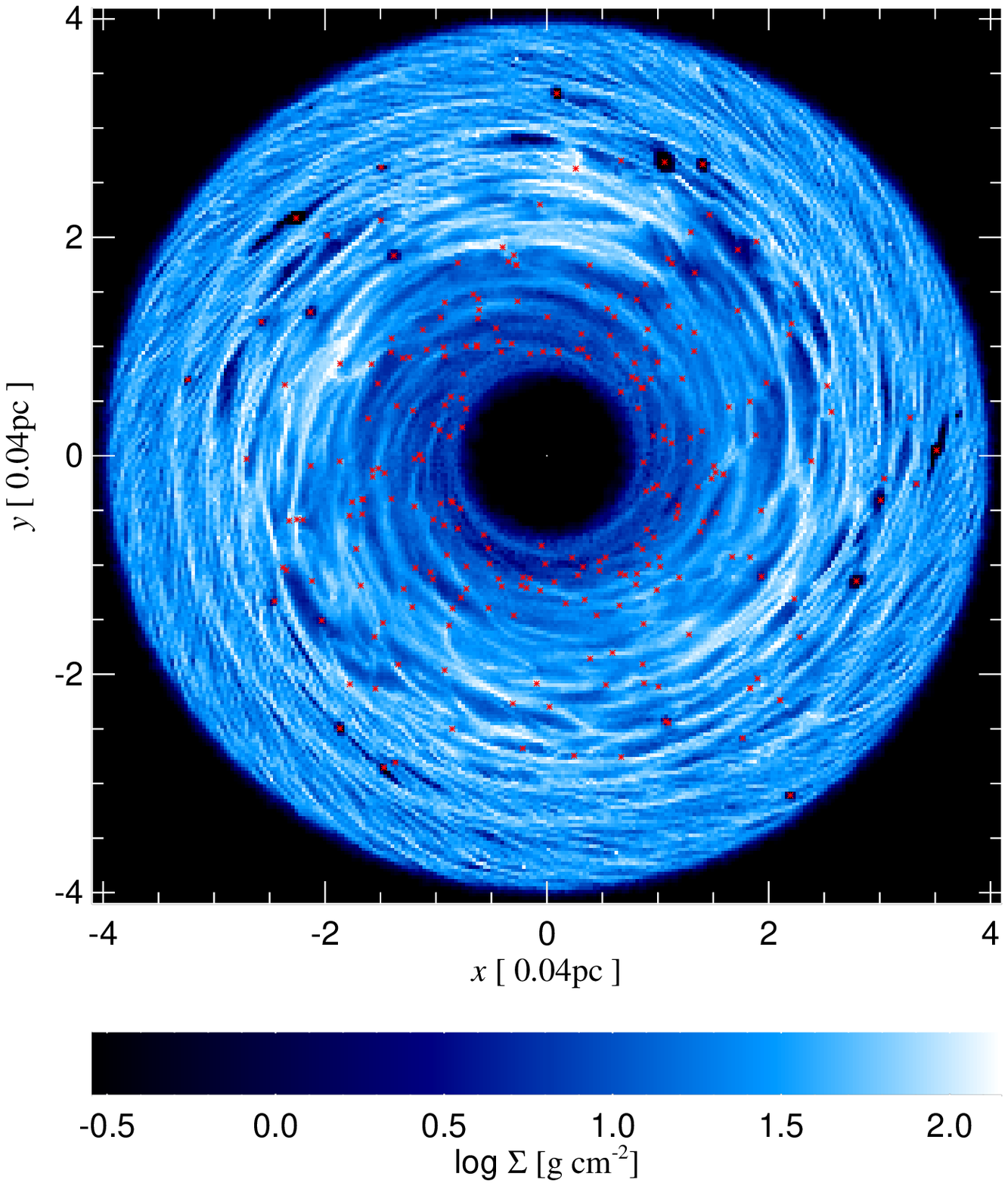,width=0.99\textwidth,angle=0}}
\end{minipage}
\caption{Snapshots of the disc for test S3 at times $t=175$ (left
  panel) and $t=275$ (right panel). Together with the left panel of
  Fig.~\ref{fig:fig2}, these snapshots trace the gradual depletion of
  the gaseous disc and the build-up of the stellar disc.}
\label{fig:fig3}
\end{figure*}

Figure~\ref{fig:fig3} shows the column density maps for the run S3 at
two later times, $t=175$ and $t=275$, in the left and right panels,
respectively. Star formation spreads to larger radii as time
progresses. In the right panel of Fig.~\ref{fig:fig3}, the innermost
region of the disc is almost completely cleared of gas by continuing
accretion onto existing stars (rather than further disc fragmentation,
see Section~\ref{sec:quenching}). At the end of the simulations that
do fragment, i.e. S1--S3, very little gas remains. To characterise the
rate at which star formation consumes the gas in the disc, we define
the disc half-time, $t_{\rm half}$, as the time from the start of the
simulation to the point at which the gaseous mass in the disc is equal
to one half of the initial mass. Table~1 lists $t_{\rm half}$ for each
simulation.  Longer cooling times allow the disc to survive
longer. Nevertheless, even for the run S3, where $t_{\rm half} \gg
t_{\rm dyn}$, the half-time of the disc is only around 12,000 years in
physical units.

One notices that some of the stars are followed and/or preceeded by low column
depth regions. These are caused by angular momentum transfer from the stars to
the surrounding gas \citep{GoldreichTremaine80}. If stars are massive enough,
they could open up gaps in accretion discs \citep{Syer91,Nayakshin04} in the
way planets open gaps in proto-planetary discs. We do not observe such
``clean'' gaps around stars in our simulations. We believe the main reason for
this is the large number of stars that we have here. The interactions between
the gravitational potentials of neighboring stars destroy the resonant
charachter of the gas-star interactions \citep{GoldreichTremaine80}. In
addition, self-interactions between stars force stars on slightly elliptical
and slightly inclined orbits \citep{NC05}, which also makes gap opening
harder.

\subsection{Evolution of disc vertical thickness}\label{sec:thickness}

The disc starts off as a razor-thin gaseous disc and evolves into a thicker
stellar disc. This behaviour is to be expected. Initially, gas cooling is
sufficiently strong, and the disc temperature is regulated to maintain a
geometrical thickness of order of $H/R \simeq M_{\rm disc}/\mbh$, which is
equivalent to having $Q\approx 1$ \citep[e.g.,][]{Gammie01}. When a good
fraction of the gas is turned into stars, however, the rate at which the disc
can cool decreases. At the same time, a stellar disc, considered on its own,
would thicken with time since stellar orbital energy is transferred into
stellar random motions by the normal N-body relaxation processes.

\begin{figure*}
\begin{minipage}[b]{.48\textwidth}
\centerline{\psfig{file=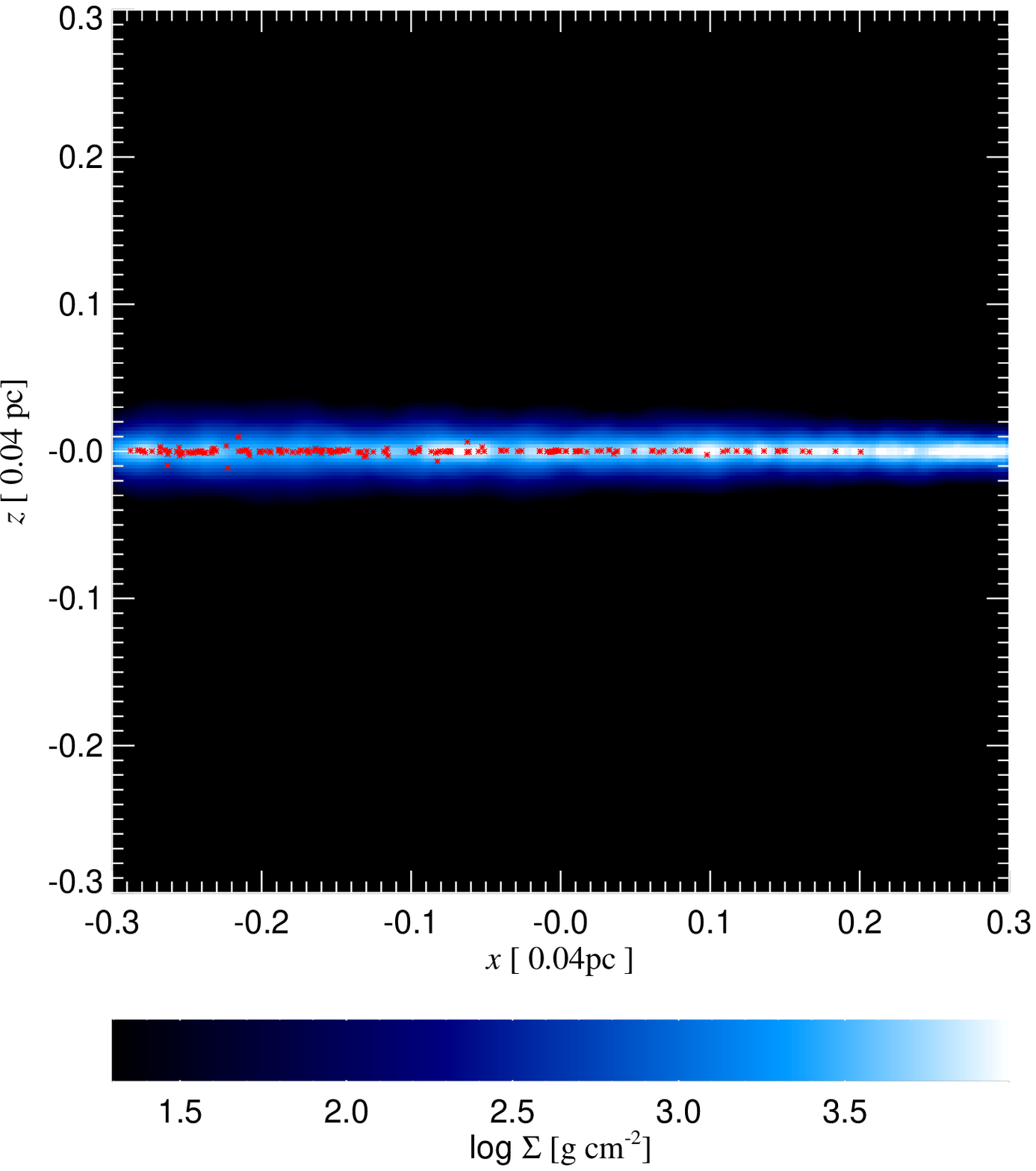,width=0.99\textwidth,angle=0}}
\end{minipage}
\begin{minipage}[b]{.48\textwidth}
\centerline{\psfig{file=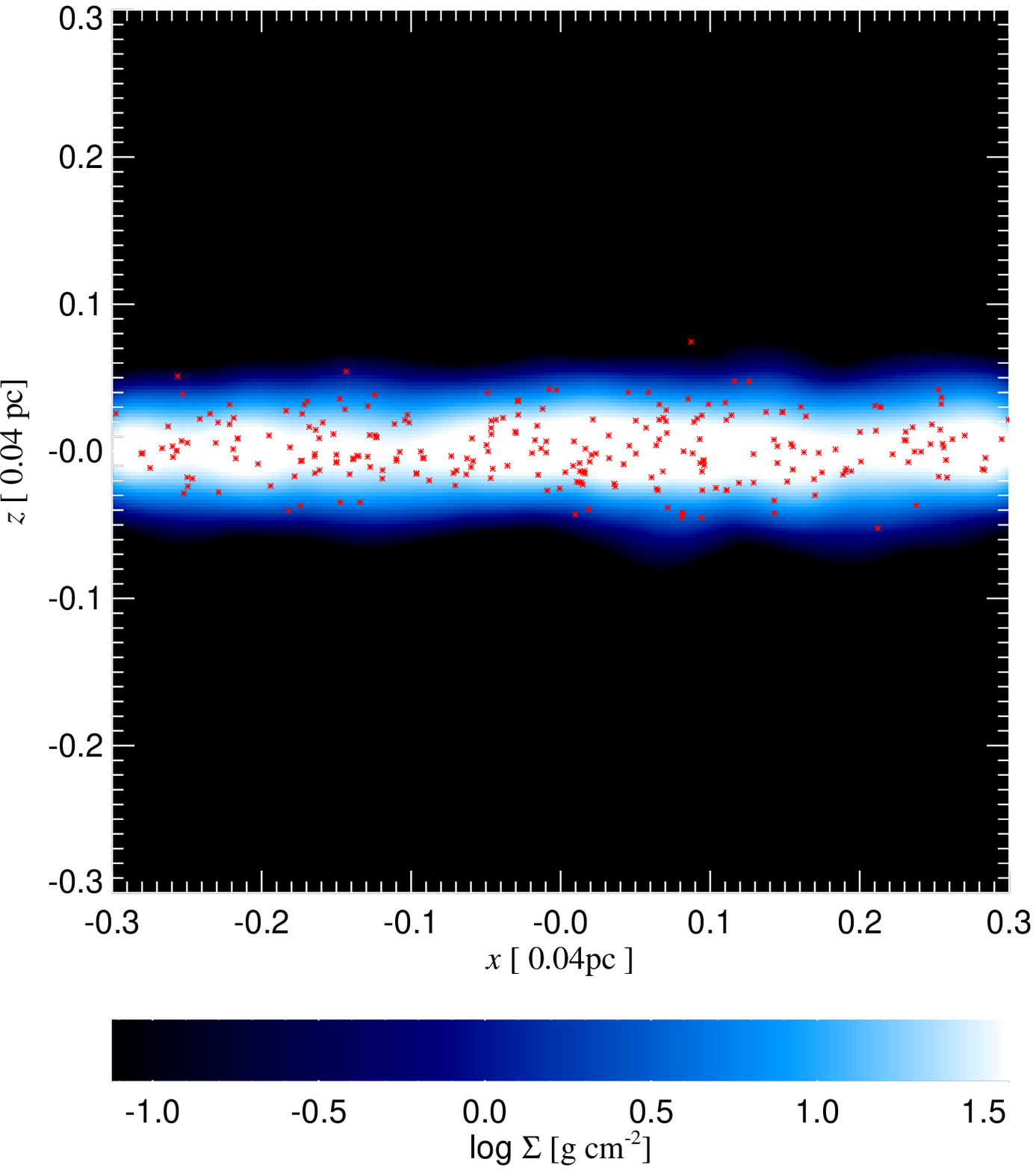,width=0.99\textwidth,angle=0}}
\end{minipage}
\caption{Evolution of the gaseous and stellar disc thicknesses in the test
  S2. The image is centred at $x=1.3$. The left panel is plotted at time
  $t=50$, when $\sim 20$\% of gas was turned into stars, whereas the right one
  is for $t=175$, at which point only $\sim 15$\% of gas remains in the
  disc. The disc thickens with time as the remaining gas is not able to
  provide sufficient damping and cooling of stellar $N$-body heating.
\label{fig:edgeon}}
\end{figure*}

In the intermediate state, when both gas and stars are present in important
fractions, disc evolution is not trivial. In the runs described in this
section, stars transfer their random energy to the gas via Chandrasekhar's
dynamical friction process \citep{Nayakshin06a}.  The disc swells as time
progresses, as can be seen in the two edge-on views of the disc from
simulation S2 shown in Fig.~\ref{fig:edgeon} . The gaseous disc can thus
become thicker than it would have been on its own. However, depending on
conditions (i.e. cooling parameter $\beta$, initial total disc mass, mass
spectrum of stars), the two discs can be coupled or decoupled. In the latter
case the stellar disc has a larger geometrical thickness than the gaseous disc. In such a case the rate at
which stars heat the disc is not trivially
calculated. This is especially so if stellar radiation, winds and supernovae
are taken into account, as stellar energy release above the disc is much less
effective in disc heating than it is inside the disc.

\subsection{Fragmentation quenching}\label{sec:quenching}

\begin{figure}
\centerline{\psfig{file=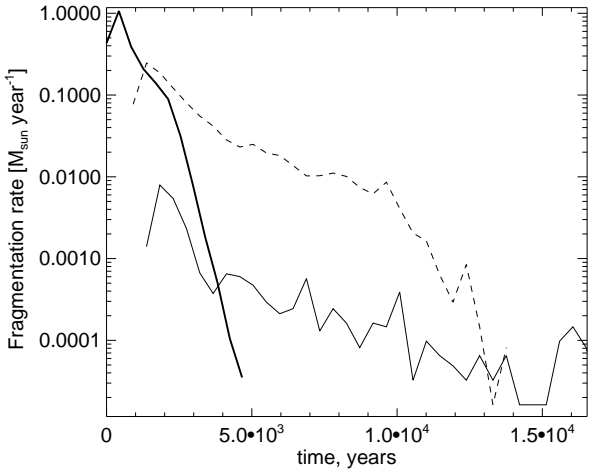,width=.48\textwidth,angle=0}}
\caption{Disc fragmentation rate in the tests with $\beta=0.3$ (S1,
  thick solid), $\beta=2$ (S2, dashed), and $\beta=3$ (S3, thin
  solid). The shorter the cooling time, the faster disc
  fragmentation proceeds.}
\label{fig:fragmrate}
\end{figure}

We expect that, following an increase in disc effective temperature
and geometrical thickness with time, the vertically averaged density
of the disc must drop. The disc may then evolve into a non
self-gravitating state as $Q$ increases above unity (see equation
\ref{q}). Further disc fragmentation should cease.  This effect can be
noticed in the right hand panel of Fig.~\ref{fig:fig1}. Near the inner
edge of the disc, there are stars but no high density clumps or
filaments implying that the disc is no longer fragmenting in that
region.

Figure~\ref{fig:fragmrate} shows the radius-integrated fragmentation rate of
the disc in the simulations with $\beta=0.3$ (run S1, thick solid curve),
$\beta=2$ (S2, dashed) and $\beta=3$ (S3, thin solid). The fragmentation rate
is defined as the total mass of first cores created per unit time. Not
surprisingly, the shorter the cooling time (smaller $\beta$), the more rapid
is disc fragmentation. The more vigorous disc fragmentation explains why there
are more stars and dense bound gas clumps in Fig.~\ref{fig:fig1} than in
Fig.~\ref{fig:fig2} at the same time ($t=75$).

One can also see that the expectation of a decrease of the fragmentation rate
with time is borne out. At the peaks of the respective curves, most of the
disc mass is still in gaseous form. Therefore, the decline in the
fragmentation rate with time is indeed mostly a consequence of a change in the
disc state (higher $Q$-parameter) rather than due to the disc running out of
gas.

\subsection{IMF of the formed stars}

\begin{figure}
\centerline{\psfig{file=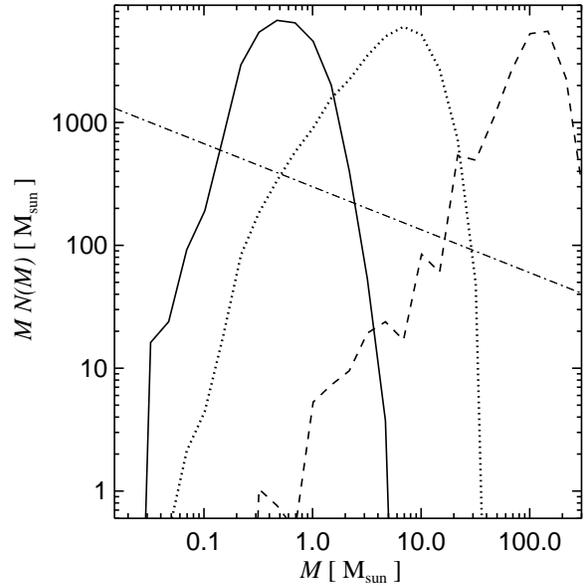,width=.48\textwidth,angle=0}}
\caption{Mass function of stars (``IMF'') formed in the simulations
  with $\beta=0.3$ (S1, solid), $\beta=2$ (S2, dotted), and
  $\beta=3$ (S3, dashed). The dot-dashed power-law in this and the
  following plots corresponds to the Salpeter IMF.}
\label{fig:imf23}
\end{figure}

It is interesting to compare the distribution functions of stellar masses from
our simulations. The end state of our simulations, i.e., when the majority of
gas is turned into stars, corresponds to the ``initial mass function'' (IMF)
of a stellar population. Therefore we use this name to refer to our mass
distributions.  Figure~\ref{fig:imf23} shows the IMF of stars formed in the
three simulations S1--S3. Table 1 lists the first two moments of the
distribution, i.e., the average mass, $\left<M_*\right>$, and
$\left<M_*^2\right>^{1/2}$. It is clear from both the figure and the table,
that the longer the cooling time, the more top-heavy (or, equivalently,
bottom-light) is the resulting IMF. This outcome is not surprising. As we saw
in Section~\ref{sec:quenching}, fragmentation is fastest for the smallest
values of the cooling parameter $\beta$. Further, fragmentation stalls in our
models when the stars heat up the disc above $Q=1$ (see
Fig.~\ref{fig:fragmrate}). At this point, disc fragmentation stops but
accretion onto stars continues. The average mass of a star reached by the time
the gas supply is exhausted is roughly inversely proportional to the number of
stars at the time the disc fragmentation stalls. As we find many more stars in
the tests where fragmentation is quickest, these simulations will also have
the least massive IMF.

The shape of stellar IMF formed in our simulations is very different from the
IMF in the Solar neighbourhood, which is close to the power-law function of
\cite{Salpeter55}. Several authors argued that this might be expected based on
theoretical grounds \citep{Morris93,Nayakshin06a,Larson06,Levin06}. The
proportion of massive stars in the stellar population of these discs might be
very important for the physics of these discs, e.g., their survival chances
against star formation.

\section{Sensitivity of results to parameter changes}\label{sec:sens}

Due to numerical limitations, our simulations do not include a number of
important physical processes. In particular, radiation transfer and chemical
processes taking place during collapse of a protostar are dealt with merely
via prescriptions. These prescriptions employ parameters which we have tried
to set to as physically realistic values as possible. It is important to
understand to which degree our results depend on the choices we made for the
values of these parameters. Here we describe a selection of tests aimed
towards this goal. These tests, labelled E1--E8 in Table~1, were done for
slightly different parameters than runs S1--S5. In particular, the inner and
outer radii were set to $r_{\rm in} =1$ and $r_{\rm out} = 2$, respectively,
and the gravitational focusing parameter was chosen to be $f_m = 1$. The
radiative cooling parameter $\beta$ was set to $3$ for all of these tests.

\subsection{Maximum accretion rate}\label{sec:mva}

We calculate the accretion rate using the Bondi-Hoyle formalism, with the
restriction that it is not allowed to exceed the Eddington limit
(equations~\ref{mbondi} and \ref{medd}). In principle, excess angular momentum
in the gas flow around proto-stars could decrease the accretion rate below our
estimates. To explore possible changes in the results, simulations E1 and E3
differed only by the maximum accretion rate allowed, with $1$ and $0.1$ times
the Eddington limit, respectively.

\begin{figure}
\centerline{\psfig{file=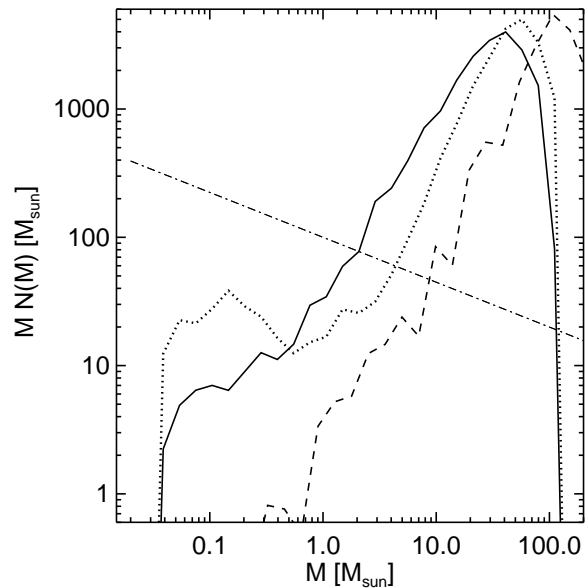,width=.48\textwidth,angle=0}}
\caption{IMF from the simulations with two different maximum allowed
accretion rates (see Section~\ref{sec:mva}), i.e., 0.1 times the Eddington
rate (E3 in Table 1; dotted curve) and the Eddington rate (E1;
solid). These curves are compared with the IMF from simulation S3
(dashed).}
\label{fig:imf_beta3}
\end{figure}

The resulting IMFs of these two runs are shown in Fig.~\ref{fig:imf_beta3},
along with the IMF for test S3, which was already presented in
Fig~\ref{fig:imf23}. Comparing the solid curve (simulation E1) with the dashed
curve (S3), we see that changes in $r_{\rm out}$, $A_{\rm col}$ and $f_m$ did
influence the IMF, making it less top-heavy. The main effect stems from the
reduction in the value of the gravitational focusing parameter, $f_m$. Fewer
mergers imply that the final average stellar mass is smaller.

Comparing the runs E1 and E3, one notices that while the IMFs of the two tests
are similar at the high mass end, around the peak in the curves, there is a
larger discrepancy at the low mass end. The largest difference in the IMFs
occurs at around the critical mass, $M_{\rm crit}=0.1\msun$. The differences
are nevertheless not as large as one could expect given an order of magnitude
change in the maximum accretion rate. We interpret these results as follows. A
star is first born as a low mass star in our simulations, and it then gains
mass by either accretion or mergers with the first cores. If the maximum
possible accretion rate onto the star is high, it ``cleans out'' its immediate
surroundings of gas, and gains most of its mass by accretion. On the other
hand, if the accretion rate is capped at lower values, as in the run E3, then
gas gravitationally captured by the star will first settle into a small scale
disc around the star. This disc is somewhat smaller than the Hill's radius
$R_{\rm H} = R (M_*/3 M_{\rm BH})^{1/3} \approx 0.01 R$ for the star of mass
$M_*=10\msun$.  The density in that disc will often exceed the critical
density for star formation, and hence new first cores will be introduced in
close proximity to the star. These cores then mostly and relatively quickly
merge with the star. Because the rate of gas capture from the larger disc
\citep[the Hill or the Bondi rate, e.g.,][]{GoodmanTan04,NC05} is controlled
by the total mass inside the Hill radius, the rate at which the dominant star
inside the Hill sphere is growing is comparable. The end result (the IMF for
massive stars) is then similar, regardless of whether the captured gas was
directly accreted or first turned into fluffy proto-stars (the first cores)
and then accreted.

Another difference between the results of the runs E1 and E3 is the
longer disc life-time, $t_{\rm half}$, for the former simulation for
rather obvious reasons.

\subsection{Critical collapse density}\label{sec:acol}

Our fragmentation criterium is controlled by two parameters, $\rho_0$ and
$A_{\rm col}$ (equation \ref{rhosink}). The default values for these
parameters are $\rho_0 = 5 \times 10^{-11}$ g~cm$^{-3}$, and $A_{\rm
col}=60$. These values were used in run E3. The simulations E2, E3a and E4 are
completely analogous to the run E3 except for the two parameters controlling
the fragmentation. In E2, the fragmentation density threshold is drastically
relaxed, with $\rho_0 = 5 \times 10^{-12}$ g~cm$^{-3}$ and $A_{\rm col}=9$.
In run E3a, $\rho_0 = 5 \times 10^{-12}$ g~cm$^{-3}$, but $A_{\rm col}=60$, as
in E3. Finally, in the run E4 we used $\rho_0 = 5 \times 10^{-11}$
g~cm$^{-3}$, and $A_{\rm col}=300$.

\begin{figure}
\centerline{\psfig{file=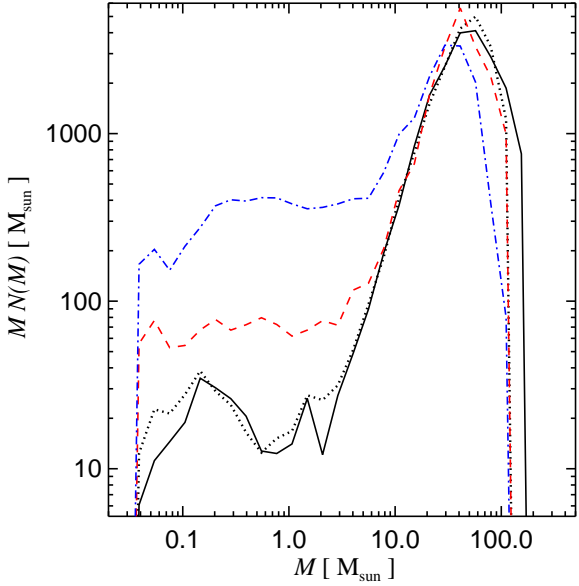,width=.48\textwidth,angle=0}}
\caption{IMF of runs E2, E3, E3a and E4 (blue dash-dotted, red dashed, dotted
  and solid curves, respectively). The simulations differ by the value of the
  critical collapse density (see Section \ref{sec:acol} for details).}
\label{fig:imf_acol}
\end{figure}

Figure~\ref{fig:imf_acol} shows the resulting mass function for the stars
formed in these four tests. Evidently, a significant relaxation of the
fragmentation criteria resulted in very large changes in the IMF at its the
low mass end. The high mass end is not so strongly affected. In fact, the IMF
of the tests E3 and E4 are almost identical, implying that the results become
independent of the value of the critical collapse density (equation
\ref{rhosink}) when it is chosen large enough.

\subsection{First core size}\label{sec:size}

\begin{figure}
\centerline{\psfig{file=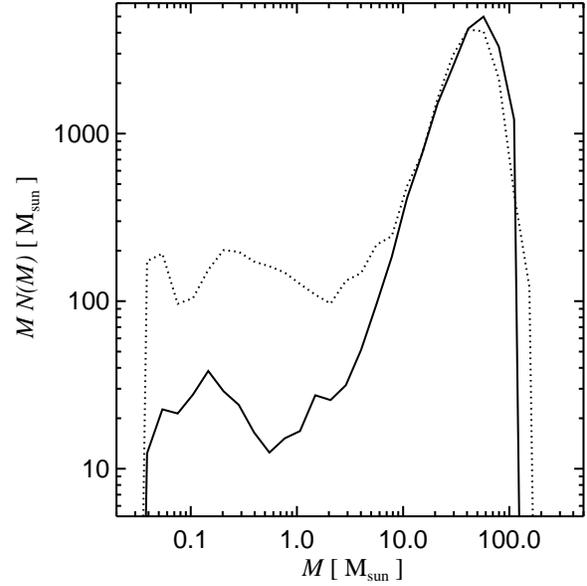,width=.48\textwidth,angle=0}}
\caption{Comparison of the IMFs of runs E3 (solid) and E5 (dotted). The
simulations have a different ``first core'' radius parameter, $R_{\rm core}=
10^{14}$~and~$3\times 10^{13}$ cm, for E3 and E5, respectively. See text in
Section~\ref{sec:size} and Table~1 for details.}
\label{fig:imf_rprot}
\end{figure}

Runs E3 and E5 are different only in the size of the first cores, $R_{\rm
core}$ (see Table~1). Figure~\ref{fig:imf_rprot} demonstrates that the IMF of
the simulations is remarkably similar at the high mass end, where it contains
almost all of the stellar mass. At the low mass end, however, simulation E5
has significantly more mass than E3.  Clearly, the smaller first core radius
$R_{\rm core}$ permits fewer mergers and slower growth of first cores by
accretion. The exact value of $R_{\rm core}$ is hence important for the
low-mass end of the IMF.

\subsection{SPH particle mass}\label{sec:mass}

\begin{figure}
\centerline{\psfig{file=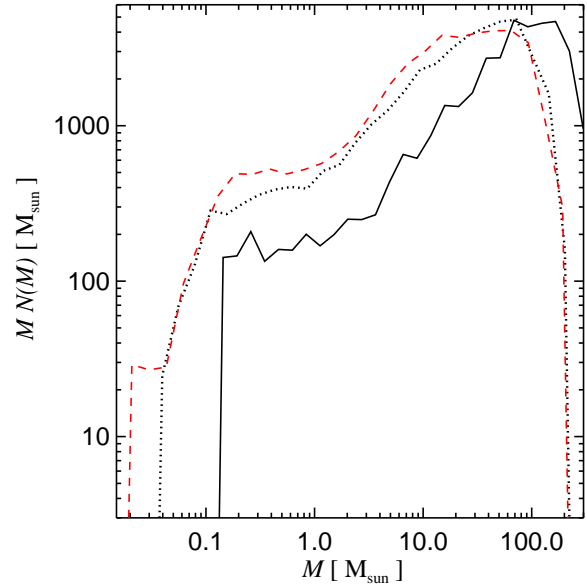,width=.48\textwidth,angle=0}}
\caption{IMF of runs E6--E8. The simulations differ in the initial
  number of SPH particles used. In particular, $N_{\rm sph} = 1$,
  2 and $4\times 10^6$ for E6, E7 and E8, respectively.}
\label{fig:imf_resolution}
\end{figure}

The simulations E6--E8 had identical parameter settings but used
different initial SPH particle numbers to test the sensitivity of the
results on the SPH mass resolution. Figure~\ref{fig:imf_resolution}
shows the IMF from these tests. Clearly, there is a significant
difference between run E6 (solid; 1 million SPH particles) and the two
others, E7 (dotted; 2 million) and E8 (dashed; 4 million). There is a
far smaller difference between the two runs with the highest
resolution, though, suggesting that the results are reasonably close
to full convergence. Given other uncertainties of the runs, e.g., the
exact value of the parameter $R_{\rm core}$ and the cooling parameter
$\beta$, etc., it seems sufficient to have an SPH mass resolution at
the level of the run E7, which has $m_{\rm sph} = 0.02\msun$. Note
that except for the simulations E6 and E7, all the other runs
presented here used $m_{\rm sph} = 0.01\msun$, as in the test E8.

\section{IMF and feedback}\label{sec:imfandfb}

\cite{Nayakshin06a} suggested that the luminosity of the young stars
which accrete gas from within a star-forming disc is sufficient to
heat up the disc, increasing the \cite{Toomre64} $Q$-parameter of the
disc above unity and hence making the disc stable to further
fragmentation. \cite{Nayakshin06a} also suggested that this will lead
to a top-heavy IMF for stars born in such marginally unstable
self-gravitating discs. To test these ideas, we ran two additional
simulations that had thermal accretion feedback included. The feedback
is implemented in the same way as in \cite{SpringelEtal05}, with a
parameter $0 \leq F\leq 1$ introduced to quantify the fraction of the
accretion luminosity that is fed back as heat into the surrounding
gas. The accretion luminosity of course depends on the accretion rate
onto the star (equation \ref{laccr}). The feedback luminosity, $F
L_{\rm acc}$, is spread over the SPH neighbours of the accreting star
according to their weight in the SPH kernel.

Our two test simulations F1 and F2 had feedback parameters of 0.01 and 0.5,
respectively, and are otherwise identical to the run E3, which had $F=0$. One
notices right away from Table~1 that the disc lifetime $t_{\rm half}$ becomes
longer for the tests with feedback compared with the test E3. In particular,
even for F1 with $F=0.01$, the disc lifetime is almost twice larger than for
E3. The test F2 in fact could not be ran until ``completion'', i.e. until most
of the gas turned into stars, as star formation was severely inhibited in that
test. At the end of the simulation, after $10^3$ time units, the disc lost
only about 10\% of its mass. We hence could only estimate that $t_{\rm half}$
will be around $10^4$ time units.  Therefore, the expectation of a stalling of
star formation is confirmed in our simulations with the large feedback
parameter $F$. However, one has to note that in a more sophisticated approach
the disc cooling time needs to be determined self-consistently, whereas we
here kept it constant artificially.

\begin{figure}
\centerline{\psfig{file=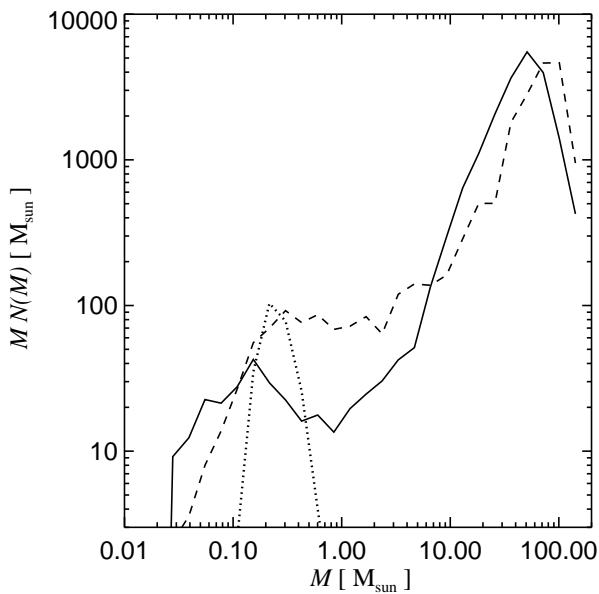,width=.48\textwidth,angle=0}}
\caption{Mass function of stars (``IMF'') formed in the simulations
 without (solid, run E3) and with feedback. The dashed curve is for
 run F1, $F=0.01$, whereas the dotted one is for F2, $F=0.5$. Note
 that in simulation F2, only $\sim 200\msun$ of stars were formed by
 the end of the simulation.  See text in Section~\ref{sec:imfandfb}
 for details.}
\label{fig:imf_fb}
\end{figure}

The IMFs of the runs F1 and F2 are compared to that of the run E3 in
Figure~\ref{fig:imf_fb}. The IMF of the low feedback parameter case, F1, is
biased towards massive stars slightly more than that of the simulation E3,
providing some support for the predictions of \citet{Nayakshin06a}. However,
simulation F2 has not been run for long enough to produce a meaningful stellar
mass function.  In perspective, we suspect that the constant cooling time
model is a poor approximation especially when stellar feedback is explored. We
shall develop a more realistic cooling model to test the influence of feedback
on the disc evolution in the future.

\section{Elliptical orbits}\label{sec:ecc}

One interesting question raised by the observations of young stars in the
central parsec of our Galaxy is that there is apparently a population of young
stars on orbits that are quite eccentric \citep[eccentricities of these stars
are of the order of 0.7; see][]{Paumard06}. The two-body relaxation time scale
is as long as $\sim 10^9$ years at this location \citep{Freitag06}, so if the
origin of these stars lies in in situ star formation, such large
eccentricities could not be built up from initial circular orbits
\citep{AlexanderBA06}; consequently their initial orbits should have been
eccentric. 

To examine whether stars can form in an eccentric gaseous disc or a stream, we
set up an additional test that is identical to simulation E1, except for the
initial conditions. In this run, designated as ``Ecc'' in Table~1, the initial
gas configuration is that of a disc segment characterized by $ 4 \le R \le 7 $
and azimuthal angle $ 0 \le \phi \le \pi/4$. Further, instead of placing the
gas on Keplerian circular orbits with $v=v_{\rm K}(R)=\Omega(R)R$, we
decreased the gas velocity to $v = 0.7 v_{\rm K}(R)$ while keeping its
direction perpendicular to its position vector and in the plane of the gas, as
in a circular disc.  These initial conditions are artificial, i.e.~unlikely
reached in a realistic infall of a gas cloud into the inner parsec of the
Galaxy. However they allow us to test star formation in an eccentric gas disc
under simple and controlled conditions.

In addition, we also introduced in this simulation a fixed spherical
potential of the older stellar cusp as deduced observationally by
\cite{Genzel03a}. The total stellar mass of this cusp within the
region being simulated is about $\sim 10$\% of the mass of \sgra,
i.e. small but not negligible.

\begin{figure*}
\begin{minipage}[b]{.4\textwidth}
\centerline{\psfig{file=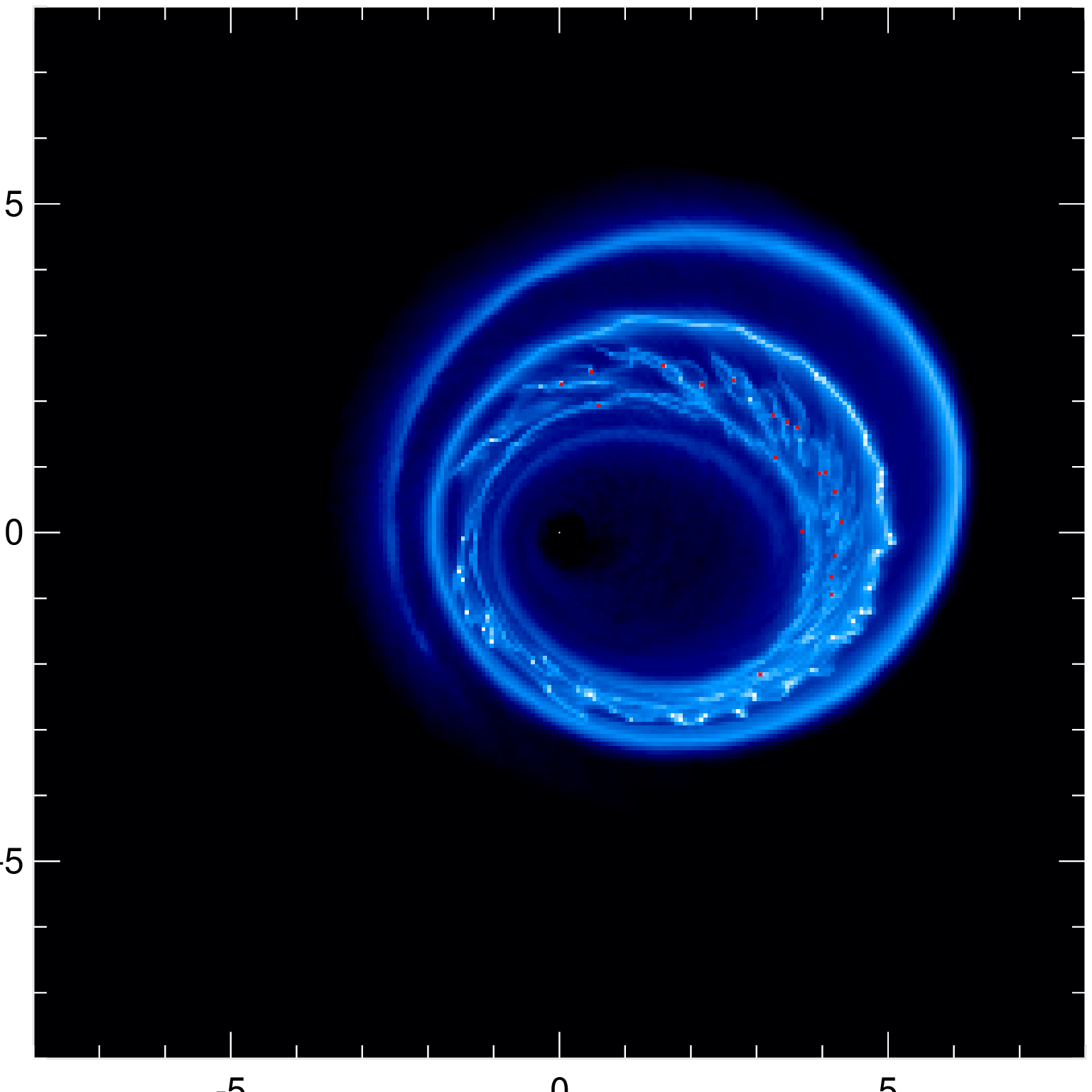,width=0.99\textwidth,angle=0}}
\end{minipage}
\begin{minipage}[b]{.4\textwidth}
\centerline{\psfig{file=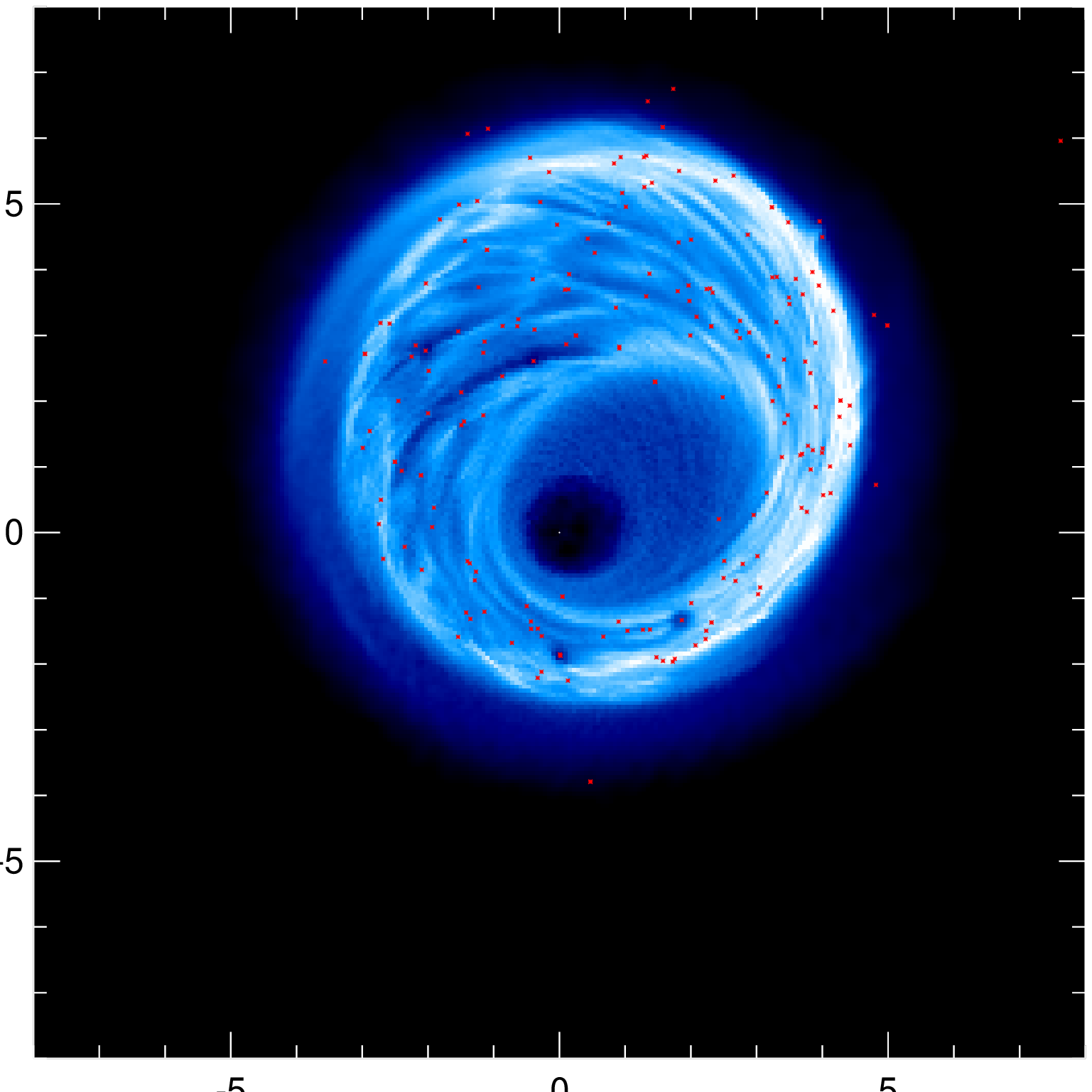,width=0.99\textwidth,angle=0}}
\end{minipage}
\begin{minipage}[b]{.4\textwidth}
\centerline{\psfig{file=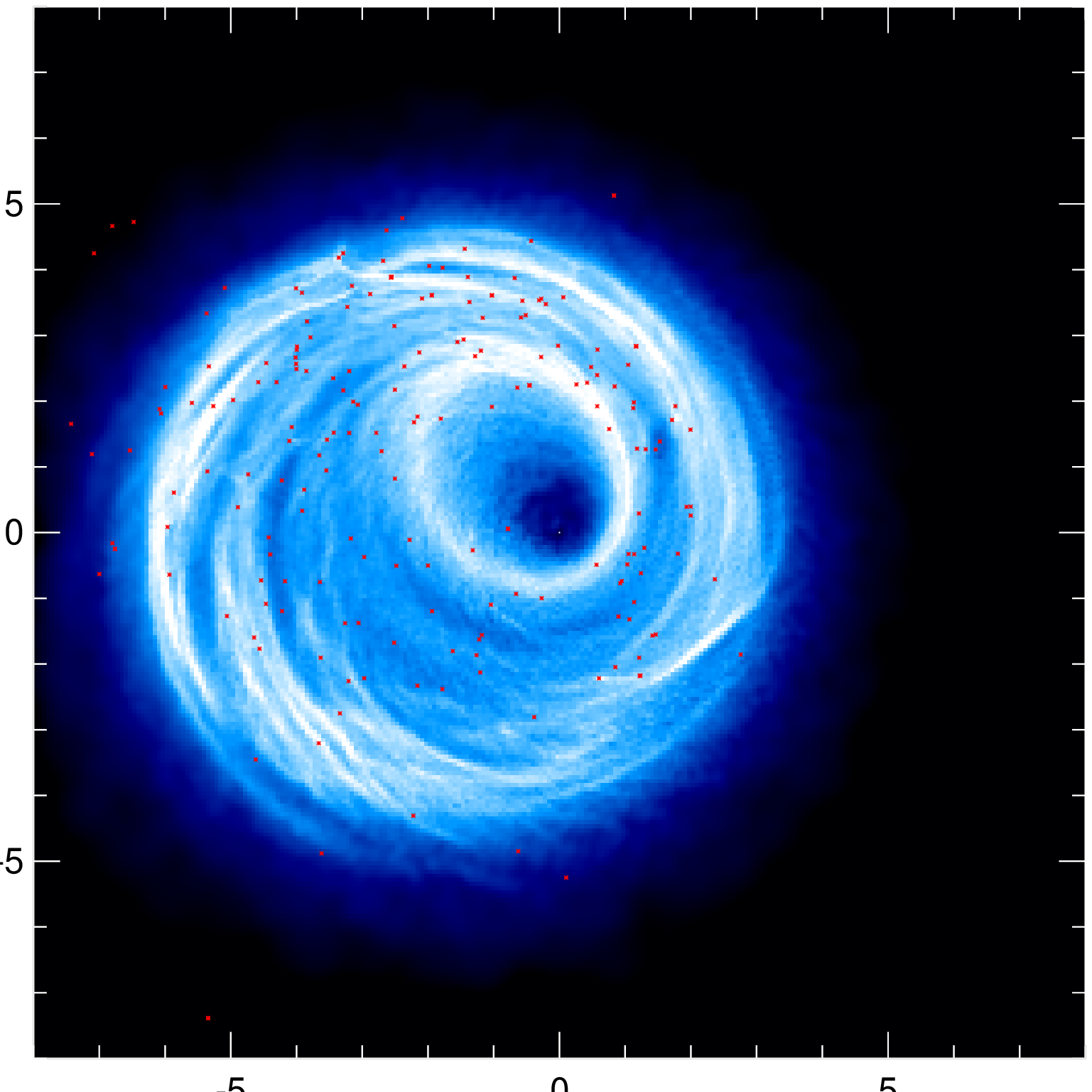,width=0.99\textwidth,angle=0}}
\end{minipage}
\begin{minipage}[b]{.4\textwidth}
\centerline{\psfig{file=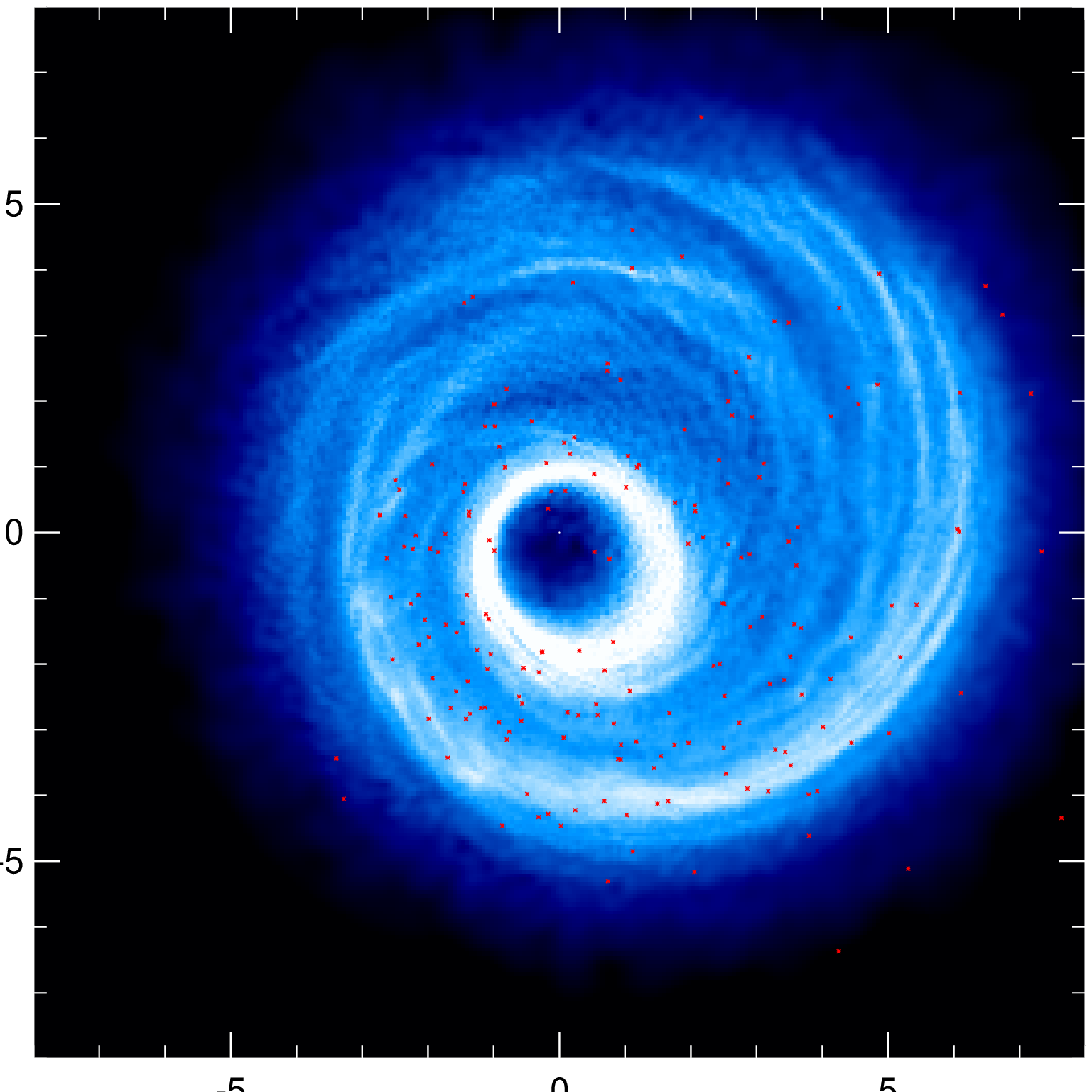,width=0.99\textwidth,angle=0}}
\end{minipage}
\caption{Evolution of the eccentric disc simulation at times
  corresponding to $t=150$, $450$, $1000$, and $2000$, from left to
  right and top to bottom. Initially, the disc segment is tidally
  sheared into an eccentric spiral. While precessing, the spiral forms
  an eccentric disc. Outer and inner edges of the disc precess at
  different rates which leads to shocks and circularization of gas
  orbits. During the simulation, the outer ring of gas makes about 30
  rotations and precesses around one full circle.}
\label{fig:ellipt}
\end{figure*}

Figure~\ref{fig:ellipt} presents several snapshots of the surface density for
test Ecc, along with stars more massive than 3~$\msun$. Initially the disc
segment is torn into an eccentric spiral that makes a few rotations around
\sgra. The spiral develops kinks (which would be essentially parts of spiral
arms in a circular gas disc) that run at an angle to the gas spiral. Stars
are born within these kinks, in orbits with eccentricities $\sim 0.5$, similar 
to the original gas orbits.
The process differs from the results of
\cite{Sanders98}, who simulated the formation of young Galactic Centre stars
on elliptical orbits in a sticky-particle approach. In that approach it was
found that star formation occured in shocks due to self-collisions of the
stream. It is possible that such a shock-mediated star formation would also
occur in our model if gas was allowed to cool much faster than the local
dynamical time.

Due to the non-Keplerian potential in this test, gas and stellar orbits
precess.  The precession rate is different at different radii, and therefore
gas orbits become mixed and somewhat circularized over time. This is most
clearly seen in the last snapshot (lower right) in Fig.~\ref{fig:ellipt},
where an inner, only mildly eccentric gaseous ring is present.

The simulation Ecc was run for twice longer than our usual
$10^3$ time units. Nevertheless, at the end of the simulation only
$\sim 30$\% of gas was turned into stars. We hence estimate the disc
half-life time $t_{\rm half}$ to be around 3000 time units. This is
$\sim 35$ times longer than that of the corresponding simulation E1
with circular orbits. While some of the difference is simply due to a
longer orbital time at the location of the gas in test Ecc, a fair
fraction of the difference is due to a comparatively slower star
formation. This is not altogether surprising, as gas is heated due to
shocks in the eccentric, precessing disc of simulation Ecc.

\begin{figure}
\centerline{\psfig{file=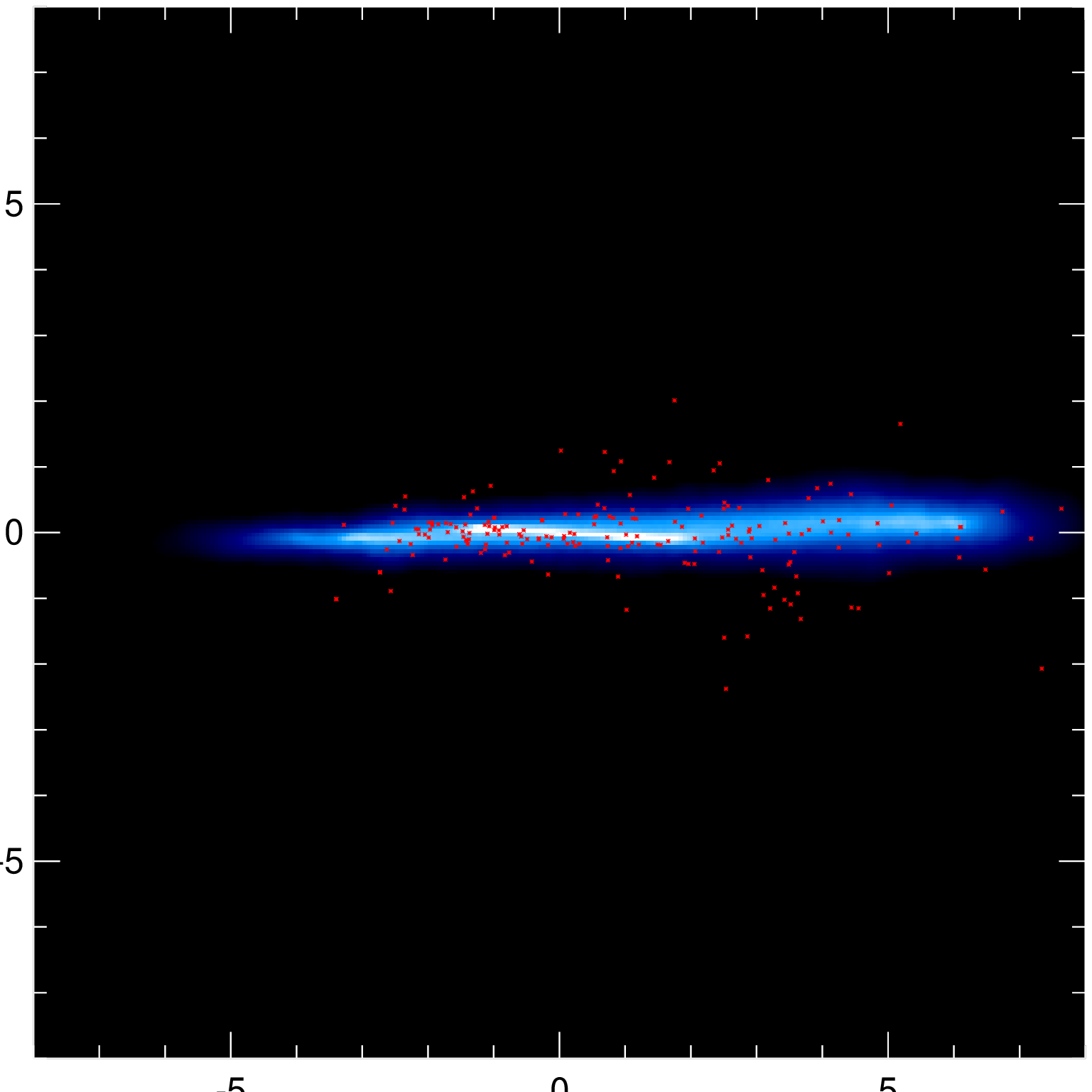,width=.48\textwidth,angle=0}}
\caption{Edge-on column density and stellar positions for the last of
  the snapshots from Fig.~\ref{fig:ellipt}. Note that the stellar disc
  is much thicker than the gaseous disc.}
\label{fig:ecc_side}
\end{figure}

Figure~\ref{fig:ecc_side} presents an edge-on view of the column
density of the disc and stellar positions at the end of the run
Ecc. It is quite noticeable that the stellar disc is much thicker than
the gaseous disc. While it is difficult to pinpoint the exact reason
for this high geometrical thickness, it is most likely due to
interactions between stars \citep[e.g.,][]{NC05,AlexanderBA06}, which
can be substantial given that this simulation has been run for over
$10^5$ years. Disruptions of gas--star clumps at their orbits'
pericentres or during collisions with each other in the disc could
also be important.

\section{in situ formation of ``mini star clusters''}\label{sec:irs13}

Several authors speculated that self-gravitating AGN discs can form
very massive objects. \cite{GoodmanTan04} argued that massive stars
formed in such discs will continue to accrete until they reach the
``isolation'' mass, $M_{\rm i}\simeq M_d^{3/2} M_{\rm BH}^{-1/2}$,
\begin{equation}
M_i \approx 550 \msun \;
\left[\frac{M_d}{10^4 \msun}\right]^{3/2} \left[\frac{3\times 10^6
\msun}{M_{\rm BH}}\right]^{1/2}\;.
\label{miso}
\end{equation}
Such a super-massive star would presumably grow via accretion from a
gas disc around it. The linear sizes of the disc would be of the order
of the Hill radius of the star, $R_{\rm H} = R (M_{\rm i}/3 M_{\rm
BH})^{1/3}$. The disc may become quite massive, and fragment
further. In this way systems more complicated than a single massive star
can be formed. It is not impossible that the compact star cluster
IRS13E, which orbits \sgra\ uncomfortably close, was formed in this
way \citep{Milosavljevic04,NC05,Levin06}.

We searched for such massive bound systems containing many stars in the
results of our simulations. While we did find many binaries of massive stars,
nothing on the scale of the isolation mass was present. In terms of physical
effects that could limit the growth of such groups, a too rapid disc
fragmentation is the most likely one. Namely, for the smaller value of the
cooling parameter $\beta$, the disc fragments quickly into too many stars that
then compete for the gas. No clear ``winners'' emerge, as can be seen from the
shape of the IMF, which rolls over very quickly at the higher mass end (see
Fig.~\ref{fig:imf23}).

For longer cooling times, i.e., $\beta \simgt 1$, disc fragmentation
is less vigorous, creating fewer stars, and leading to a more
top-heavy IMF.  In this case, while the gaseous mass is still
comparable to the stellar mass in the simulations, we did find many
binaries and multiple systems, some containing up to 6 stars. However,
closer to the end of these simulations, when the gas supply was
exhausted, only tightly bound binaries remained. There seem to be two
reasons for this. Firstly, while the system is still gas-rich,
collisions between ``mini star clusters'' take place. The collisions
tend to unbind the systems (since $\beta > 1 $). A few examples of
this could be seen directly in videos of the time evolution of the
simulations produced from the snapshots. Secondly, when most of the gas
is exhausted, individual multiple systems are likely to evaporate the
less massive components in favour of tightening the most massive
binary.

These results could in principle simply mean that we do not have a sufficient
time resolution to integrate small scale dynamics of stars in these mini star
clusters. To check this, we have ran an additional simulation, identical to
the run S2, but in which the time step criterium was 2 times more stringent.
In this simulation, not presented in the Table 1, individual particle time
steps were on average half of what they were in S2. The result was very
similar in terms of the IMF (i.e. the average stellar mass and the average
stellar mass squared were different by $\sim 3$\% and $10$\% only). However, we
did find a $\sim 50\%$ increase in the number of stars with several close
neighboors. However, again, no group was even remotely close to the isolation
mass scale. The largest group contained four neighboors. While this issue
requires further numerical work, we believe that the absense of massive
densely packed stellar groups is not a numerical artefact.

Nevertheless, independently of the precision with which the numerical
integration is performed, the results might also strongly depend on the gas
cooling physics.  A more careful treatment of gas cooling physics is needed in
order to test the model of in situ formation of IRS13E more thoroughly.

\section{Conclusions} \label{sec:conclusions}

In this paper, we presented some of our first attempts to simulate
star formation in an accretion disc around \sgra in the low disc mass
case. Since the relevant physics is only weakly dependent on the mass
of the SMBH \citep{Goodman03}, we expect that the results are relevant
to more massive SMBHs as well. On the technical side, in our numerical
method we used a locally constant cooling time prescription, and we
allowed for a finite collapse time of stellar ``first cores'' and
their mergers (Sections \ref{sec:methods} and
\ref{sec:starformation}). We limited the accretion rate of stars to a
fraction of the Eddington limit. Our simulations verified that the
code reproduces disc fragmentation in agreement with previous
analytical and numerical work (Section~\ref{sec:fragm}). We also
tested the dependence of the results on some of the plausible
parameter choices (Section~\ref{sec:sens} and Table~1).

The scientific results of our work can be summarized as follows:
\begin{itemize}
\item[(1)] Qualitatively, star formation in marginally
  self-gravitating discs, i.e. those that have cooling times just
  short enough to fragment, results in a top-heavy IMF. This is due to
  the suppression of fragmentation by disc heating either through
  N-body effects or feedback from the accretion luminosity.  Mergers
  between first cores and gas accretion lead to a rapid growth of
  existing stars. Quantitatively, the IMF of our simulations does
  depend on physics that we do not completely resolve and which we model
  through several free parameters (see Table~1).

\item[(2)] Without feedback, star formation is a very rapid process
  (unless cooling is inefficient, of course). As gas is bound to
  \sgra, nothing escapes from the star forming region, and essentially
  all of the gas in the disc is turned into stars.

\item[(3)] Due to point 2, our models completely fail to feed the
  central AGN.  Even the longest ``living'' gas disc in our runs (test
  F2 with feedback) would all be drained into young stars after about
  $10^6$ years, which is far too short compared with the normal
  timescale of feeding \sgra\ through viscous angular momentum
  transfer \citep[e.g.,][]{NC05}. Modelling of higher mass discs is
  needed to establish whether star formation feedback in such discs
  could hold off the disc's demise due to star formation in favour of
  AGN accretion.

\item[(4)] Discs more massive than marginally self-gravitating discs can be
  modelled in our approach with a small cooling time parameter, $\beta\ll
  1$. Without stellar feedback, such discs undergo a very rapid gravitational
  collapse. This leads to an IMF dominated by low mass stars. See point 5
  below, however.

\item[(5)] We have explored two limiting cases of marginally self-gravitating
  discs with thermal feedback. Putting even a small fraction of the stellar
  accretion luminosity back into the disc significantly increases the disc
  lifetime, that is, it slows down star formation. A more sophisticated
  numerical treatment of heating/cooling is necessary to test whether the IMF
  of such discs will become top heavy, as predicted.

\item[(6)] We also simulated star formation in an eccentric stream of gas in
  the gravitational potential of \sgra\ and the surrounding stellar cusp. The
  gas stream precesses, self-collides and forms an eccentric disc. Massive
  stars in eccentric orbits are readily formed in the simulation. 
  Tidal shearing and shocks in the
  eccentric case do not prevent collapse, though they do increase the lifetime
  of the gaseous disc.

\item[(7)] None of the simulations formed an object similar to IRS13E, a bound
``mini star cluster'' in the GC. While more careful modelling is needed to
confirm the result, it is suggestive of a formation mechanism for this object
different from the in situ star formation in the disc.

\end{itemize}

We thank Giuseppe Lodato for detailed comments on the paper.

\begin{table*}
\begin{center}
\begin{tabular}{|l|c|c|c|c|c|c|c|c|c|c|c|c|}\hline
Simulation  & $\beta^{\, a}$ & $F^{\, b}$ & $N_{\rm sph}$, & Disc radial & $M_{\rm disc}$, 
& $A_{\rm col}^{\, c}$ & Max. accre- & $R_{\rm core}$ & $f_{\rm m}^{\, e}$ & $t_{\rm half}^f$ & $M_*$, 
& $\sqrt{M_*^2}$, \\ 
name and purpose & & & $10^6$ & extent & $\msun$ & & tion rate$^d$ &
$10^{14}$~cm & & &$\msun$ & $\msun$  \\
\hline \hline
fragmentation study\\
S1 & 0.3 & 0 & 4 & $1-4$ & $3\times 10^4$ & 30 & 1 & 1 & 3 & $10.5$ & $0.35$ & $0.43$ \\
S2 & 2. & 0 & 4 & $1-4$ & $3\times 10^4$ & 30 & 1 & 1 & 3 & $88$ & $2.6$ & $3.7$\\
S3 & 3. & 0 & 4 & $1-4$ & $3\times 10^4$ & 30 & 1 & 1 & 3 & $230$ & $59.6$ & $76.3$ \\
S4 & 4.5 & 0 & 4 & $1-4$ & $3\times 10^4$ & 30 & 1 & 1 & 3  & -- & -- & -- \\
S5 & 6. & 0 & 4 & $1-4$ & $3\times 10^4$ & 30 & 1 & 1 & 3 &  -- & -- & -- \\
\hline
Maximum $\dot M_*$\\
E1 & 3. & 0 & 2 & $1-2$ & $2\times 10^4$ & 60 & 1 & 1 & 1 & 86  & 9.9 & 17.0
\\
E3 & 3. & 0 & 2 & $1-2$ & $2\times 10^4$ & 60 & 0.1 & 1 & 1 & 240 & 5.7 & 15.5
\\
\hline
Critical density\\
E2 & 3. & 0 & 2 & $1-2$ & $2\times 10^4$ & 9 & 0.1 & 1 & 1 & 235 & 0.55 & 3.3 \\
E3 & 3. & 0 & 2 & $1-2$ & $2\times 10^4$ & 60 & 0.1 & 1 & 1 & 240 & 5.7 & 15.5
\\
E3a & 3. & 0 & 2 & $1-2$ & $2\times 10^4$ & 60 & 0.1 & 1 & 1 & 240 & 1.9 & 8.3 \\
E4 & 3. & 0 & 2 & $1-2$ & $2\times 10^4$ & 300 & 0.1 & 1 & 1 & 245 & 8.0 &
19.2 \\
\hline
First core size\\
E3 & 3. & 0 & 2 & $1-2$ & $2\times 10^4$ & 60 & 0.1 & 1 & 1 & 240 & 5.7 & 15.5 \\
E5 & 3. & 0 & 2 & $1-2$ & $2\times 10^4$ & 60 & 0.1 & 0.3 & 1 & 240 & 0.6 &
4.7 \\
\hline
Resolution study\\
E6 & 3. & 0 & 1 & $1-2$ & $4\times 10^4$ & 60 & 1 & 1 & 1 & 43 & 5.3 & 20.4 \\
E7 & 3. & 0 & 2 & $1-2$ & $4\times 10^4$ & 60 & 1 & 1 & 1 & 44 & 2.0 & 8.1 \\
E8 & 3. & 0 & 4 & $1-2$ & $4\times 10^4$ & 60 & 1 & 1 & 1 & 44 & 1.5 & 6.5 \\
\hline
feedback importance\\
F1 & 3. & 0.01 & 2 & $1-2$ & $2\times 10^4$ & 60 & 0.1 & 1 & 1 & $400$ & 5.4 & 17.4 \\
F2 & 3. & 0.5 & 2 & $1-2$ & $2\times 10^4$ & 60 & 0.1 & 1 & 1 & $\sim 10^4$ & (?) & (?) \\
\hline
Eccentric orbits\\
Ecc & 3. & 0 & 2 & $\sim 1-7$ & $2\times 10^4$ & 60 & 1 & 1 & 1 & $\approx 3000$ & 9.9 & 17.0 \\
\hline
\end{tabular}
\end{center}
\caption{Description of simulations performed in the paper. Some of the run
  are printed several times in the table to allow an easier comparison
  between the tests grouped together. Notes:} 
  \leftline{$^a$ Cooling parameter introduced in
  Section~\ref{sec:cooling}.}  \leftline{$^b$ Feedback parameter, the
  fraction of stellar accretion energy fed back into the disc
  (Section~\ref{sec:imfandfb}).}  \leftline{$^c$ Collapse over-density
  parameter (Section~\ref{sec:collapse})} \leftline{$^d$ The maximum
  accretion rate of sink particles in units of their Eddington
  accretion rate (see Sections \ref{sec:mva} \& \ref{sec:medd}).}  \leftline{$^e$ Gravitational
  focusing parameter for mergers of first cores (Section~\ref{sec:sink})}
\leftline{$^f$ Time scale for the gaseous mass in the disc to drop by half.}
\end{table*}

\bibliographystyle{mnras} 

\end{document}